\documentclass[11pt]{article}
\usepackage{amsfonts,amssymb,amstext,epsfig}
\usepackage[latin1]{inputenc}
\usepackage[english]{babel}
\usepackage{color}
\usepackage[T1]{fontenc}
\usepackage{makeidx}
\usepackage{graphicx}
\usepackage{mathtools}
\usepackage{braket}
\usepackage{}
\usepackage{amsthm}
\usepackage{systeme}
\usepackage{calrsfs}
\usepackage{hyperref}
\newtheorem{thm}{Theorem}[section]
\newtheorem{cor}[thm]{Corollaire}
\newtheorem{lem}[thm]{Lemme}
\newtheorem{pro}[thm]{Proposition}
\newtheorem{dfn}[thm]{Definition}
\newtheorem{rmk}[thm]{Remark}
\newtheorem{expl}[thm]{Exemple}

\newcommand{\ba}{\begin{array}}
\newcommand{\ea}{\end{array}}
\newcommand{\MeijerG}[5]{G^{#1}_{#2}
\def\deq{\stackrel{\mathrm{def}}{=}}
	\left(#5\middle\vert\
	\begin{smallmatrix} #3 \\ \\ \\ #4	
	\end{smallmatrix} 
	\right) }

\def\dessous#1\sous#2{\mathrel{\mathop{\kern0pt#2}\limits_{#1}}}
\def \dis {\displaystyle}

\newcommand{\C}{\mathbb C}
\newcommand{\N}{\mathbb N}

\newcommand{\beq}{\begin{eqnarray}}
\newcommand{\eeq}{\end{eqnarray}}
\newcommand{\bpro}{\begin{pro}}
\newcommand{\epro}{\end{pro}}
\newcommand{\blem}{\begin{lem}}
\newcommand{\elem}{\end{lem}}
\newcommand{\bdfn}{\begin{dfn}}
\newcommand{\edfn}{\end{dfn}}
\newcommand{\bcor}{\begin{cor}}
\newcommand{\ecor}{\end{cor}}
\newcommand{\bthm}{\begin{thm}}
\newcommand{\ethm}{\end{thm}}
\newcommand{\bex}{\begin{expl}}
\newcommand{\eex}{\end{expl}}
\newcommand{\brmk}{\begin{rmk}}
\newcommand{\ermk}{\end{rmk}}
\newcommand{\benum}{\begin{enumerate}}
\newcommand{\eenum}{\end{enumerate}}
\newcommand{\bitem}{\begin{itemize}}
\newcommand{\eitem}{\end{itemize}}
\begin{document}
   \begin{center}
 {{\Large\bf 
%        Generalized hypergeometric coherent states associated to orthogonal polynomials: algebraic and physical properties \\
       Generalized hypergeometric coherent states for special functions: mathematical and physical properties

%        On some special functions associated generalized hypergeometric coherent states and related properties 
%        
%        
%       DENSITY OPERATOR FORMULATION FOR A SUPERSYMMETRIC
% HARMONIC OSCILLATOR: VECTOR COHERENT STATE
% CONSTRUCTION AND STATISTICAL PROPERTIES 
}}

Isiaka Aremua$^{1,2}$,  Messan M\'{e}dard  Akouetegan$^{1}$ , Komi Sodoga$^{1,2}$, Mahouton Norbert Hounkonnou$^{3}$, and  Yaogan Mensah$^{4,2}$

 $^{1}${\em
Universit\'{e} de Lom\'{e} (UL), Facult\'{e} Des Sciences  (FDS), D\'{e}partement de Physique} \\
{\em Laboratoire de Physique des Mat\'eriaux et des Composants \`a Semi-Conducteurs}\\
{\em Universit\'{e} de Lom\'{e} (UL), 01 B.P. 1515  Lom\'{e} 01, Togo.}\\  
$^{2}${\em
International Chair of Mathematical Physics and Applications,
ICMPA-UNESCO Chair, University of Abomey-Calavi, 072 B.P. 50 Cotonou, Republic of Benin.}\\
$^{3}${\em  International Chair in Mathematical Physics
and Applications} \\
{\em ICMPA-UNESCO Chair, University of Abomey-Calavi,  072 B.P. 50 Cotonou, Republic of Benin}
\\ {\em and Centre International de Recherches et d'Etude Avanc\'{e}es en Sciences Math\'{e}matiques}\\  {\em \& Informatiques et Applications 
(CIREASMIA), 072 B.P. 50 Cotonou, Republic of Benin} \\ 
$^{4}${\em
Departement  of Mathematics,  
University of Lom\'{e}, Lom\'{e}, Togo. }\\
{\em   E-mails:claudisak@gmail.com, 
akoueteganmessanmedard@gmail.com, antoinekomisodoga@gmail.com, mensahyaogan2@gmail.com, 
norbert.hounkonnou@cipma.uac.bj}

\vspace{1.0cm}

\today

         %\clearpage
          \begin{abstract}
\noindent
In continuation of our previous works 
 [{\it J. Phys. A: Math. Gen.} {\bf 35}, 9355-9365 (2002)], [{\it J. Phys. A: Math. Gen.} {\bf 38}, 7851 (2005)] and 
%   has been  achieved in
    [{\it Eur. Phys. J. D} {\bf 72}, 172
(2018)], 
 we investigate a class of  generalized  coherent
states for associated Jacobi polynomials
and hypergeometric functions,    
satisfying the 
resolution of the
identity  
with respect to a weight function expressed
in terms of Meijer's G-function. 
 We extend the state Hilbert space  of the constructed states and discuss  the property of the reproducing kernel  and its analytical expansion. Further, we provide the expectation values of observables  relevant to this quantum model. 
We also perform the quantization of the complex plane, 
 compute and analyze the probability density and the temporal stability in these states. %{\color{blue}
 	Using the completeness relation provided by the coherent
states, we achieve the thermodynamic analysis in the diagonal $P$-representation of
the density operator.

\end{abstract}

\end{center}

\setcounter{footnote}{0}

\section{Introduction}

During the past few decades, the concept of coherent states (CSs) has aroused great
scientific interest since their introduction, at the beginning of the 1960s, by Schr\"{o}dinger \cite{schroedinger} for the quantum
harmonic oscillator (HO)as a specific quantum states which have the most similar dynamical behavior to that of classical HO.

% as the specific quantum state which  has dynamical behavior that is most similar to that of the classical HO.
  Glauber \cite{glauber} and Sudarshan \cite{sudarshan} reconsidered the definition
of  these 
%introduced by
 Schr\"{o}dinger CSs, 
% at the beginning of the 1960s,
  while the conditions any state must fulfill to be
coherent, (i. e., continuity in complex label, normalization,
non orthogonality, unity operator resolution with unique positive weight function of the
integration measure, temporal stability and action identity),   were elaborated by Klauder \cite{klauder}. More details on the CSs and
their different generalizations can be found in the literature. See, for example,  \cite{klauder-skagerstam,perelomov,gazeaubook,ali-antoine-gazeau}. 
% In addition, Klauder \cite{klauder} formulated
% the so-called \textquotedblleft Klauder's minimal prescriptions (or conditions) \textquotedblright     that any CS must meet: continuity in complex label, normalization, non orthogonality, unity operator resolution with unique positive weight function
% of the integration measure,   temporal stability and action identity. 
% [J. R. Klauder, J. Math.
% Phys. 4, 1055 (1963).]. 
% After nearly half a century,
% scientific interest for CSs has been returned, due to their potential applications in modern
% quantum optics, solid state physics,  mathematical physics to signal theory, quantum information, cosmology, etc \cite{klauder-skagerstam, ali-antoine-gazeau, gazeaubook} 
% [klauder-skagerstam, ali-antoine-gazeau, gazeaubook, dodonov]. 
%Since the HO,
Different kinds of CSs have also been generalized  for 
%many 
quantum systems. One can mention  
% some new classes of generalizations of CS, such as
the  Barut-Girardello CSs \cite{barut-girardello}, Perelomov CSs \cite{perelomov}, Gazeau-Klauder CSs  \cite{gazeau-klauder1}, Penson-Solomon CSs \cite{penson-solomon}
Klauder-Penson-Sixdeniers CSs \cite{klauder-penson-sixdeniers}, generalized hypergeometric CSs (GHCSs)  introduced by Appl and Schiller \cite{appl-schiller}. These have been taken into account in many works since their introduction. One can also  consult \cite{dodonov} concerning nonclassical properties of CSs. Generalized hypergeometric photon-added and photon-depleted CSs, and deformed photon-added non-linear CSs were introduced, respectively, in \cite{hounk-ngompe}
and \cite{safaeian-tavassoly}. Note that the photon-added CSs (PACSs)  and the lower truncated CSs are the limiting cases of  suitably deformed PACSs \cite{sivakumar}.  Photon-added Gazeau-Klauder and Klauder-Perelomov
 CSs for exactly solvable Hamiltonians were studied
in \cite{daoud}, while  photon-added Barut-Girardello  CSs of
the pseudoharmonic oscillator were constructed in \cite{popov1}. Besides, in \cite{popov1} the GHCSs were extended to mixed (thermal) states and applied, particularly, to the case of a pseudoharmonic oscillator.

% [V. V. Dodonov, J. Opt. B: Quant. Semiclass. Optics 4, R1
% (2002)]. 
% {\color{red}
% The required criteria for any state to be coherent
% were elaborated by Klauder: continuity in complex label, normalization, non orthogonality, unity operator resolution with unique positive weight function of the integration measure, temporal stability and action identity [J. R. Klauder, J. Math. Phys. 4, 1055 (1963).]}

% A great interest has been devoted to the GH-CSs  since their introduction by Appl and Schiller \cite{appl-schiller}. 
% They have also appeared the so called squeezed states,
% as well as the CSs with photon added. For more details on the different kind of these states,  see 
%  \cite{dodonov}.
 In general, the GHCSs are given by the expression \cite{appl-schiller,popov1}:
\beq\label{applschill}
|z\rangle  = \frac{1}{\sqrt{_pF_q(\{a_i\}^p_1;\{b_j\}^q_1;|z|^2)}}\sum_{n = 0}^{\infty}\frac{z^n}{\sqrt{\rho_{p, q}(n)}}|n\rangle,
\eeq
where $_pF_q(\{a_i\}^p_1;\{b_j\}^q_1;|z|^2)$ are the generalized hypergeometric functions \cite{mathai}:
{\small
\beq
_pF_q(\{a_i\}^p_1;\{b_j\}^q_1;x) = \sum_{n=0}^{\infty}\frac{\displaystyle\prod_{i=1}^{p}(a_i)_{n}}{\displaystyle\prod_{j=1}^{p}(b_j)_{n}}\frac{x^n}{n  
!}\equiv \sum_{n=0}^{\infty} \frac{1}{\rho_{p, q}(n)}x^n,
\eeq
}
 $p$ and $q$ are integer numbers; the $\rho_{p, q}(n)$ stand for generalized
factorials expressed through Pochhammer symbols $(a)_n = \frac{\Gamma(n+a)}{\Gamma(a)}$ and Euler gamma functions $\Gamma(x)$. The appellation of {\it generalized hypergeometric coherent state}         
       refers to the normalization function  given by generalized hypergeometric functions.

Cotfas, in his work \cite{cotfas}, provided a factorization method
of associated hypergeometric operators, and deduced the
associated algebra and corresponding CSs. These are
eigenstates of the annihilation operator denoted $a_m$. Following  Aleixo  {\it et al.} \cite{aleixoetal}, we introduced a right inverse operator  $a^{-1}_m$ 
%was introduced in a previous paper
in \cite{hounk-sodoga} in order to  define generalized associated hypergeometric CSs
(GAH-CSs). 
These states fulfill the Klauder prescriptions required for a set of CSs and establish a close connection between the quantum and classical formulations of a given physical system.
 Recently \cite{aremuaamsj}, some of us applied the general procedure of CS quantization, (also
known as the Berezin-Klauder-Toeplitz quantization), 
%was applied 
in the complex plane to a set of GPAH-CSs from the resolution of the identity   obtained by a positive
weight function expressed in terms of Meijer's G-functions. In addition, these authors discussed the nonclassical behaviour %was discussed
 by investigating the Mandel Q-parameter expressed in terms of generalized hypergeometric functions.
All these developments on  generalizations of GHCSs and GPAH-CSs,
which also put in light the applications of
hypergemetric functions and Meijer's G-functions  (see \cite{appl-schiller,cotfas,hounk-sodoga,popov1,hounk-ngompe,epjdhyper}, and references  therein), 
motivate the present work. 
Indeed, following the method developed in \cite{cotfas}-\cite{hounk-sodoga}, we built a family of generalized hypergeometric CSs (GHCSs) for associated
Jacobi polynomials and hypergeometric functions. These  CSs 
coincide with  the GHCSs introduced by Appl and Schiller \cite{appl-schiller}, (see (\ref{applschill}) in this paper), and  their expansion is provided in terms of the Fock basis states.  
All computations, including the weight function of the integration measure, 
%in the rest of the paper,
 are performed
in terms of the Meijer's G-functions related to the hypergeometric functions,   and
% also in terms of
  modified
Bessel functions of the first kind, 
% This fact
 highlighting the relevance of the usefulness of these functions.
  %  for these functions 

%and shows a certain  uniformity  of the obtained results throughout this work.

%{\color{blue}
	The diagonal expansion of
the density operator, known as the Glauber-Sudarshan (GS)-$P$-representation \cite{brif-aryeh}, plays a key role in the
concepts of Husimi distribution \cite{husimi} and Wehrl entropy \cite{wehrl} in   various constructions
 (see for example \cite{popov1,epjdhyper,popov3,a-hkn-ez-jnmp,curilef-plastino} for more details). Also in \cite{a-hk-ez-jmp}, the density operator diagonal representation in the CSs basis was used to
study harmonic oscillator quantum systems and models of spinless electrons moving
in a two-dimensional noncommutative space, subject to a magnetic field coupled
with a harmonic oscillator. Relevant statistical properties such as the $Q$-Husimi
distribution and the Wehrl entropy were also investigated. 
 Besides, multi-matrix vector coherent states (VCSs)
basis was successfully performed  in the density operator representation and applied to Landau levels of an electron in an
electromagnetic field coupled to an isotropic harmonic potential \cite{a-hk-sod-romp}. Main relevant
statistical properties such as the Mandel $Q$-parameter and the signal-to-quantum-noise
ratio were derived and discussed. In addition, more  recently \cite{a-hk-sod-tchak-romp}, in the context of supersymmetric harmonic oscillator, 
a matrix formulation of the density operator to construct a two-component VCS representation was achieved. Relevant statistical properties were described, with  a link with quantum information  given
 via an integral representation of a qubit.
%}

The paper is organized as follows. Section 2 recalls the construction method of hypergeometric CSs as it is carried out in the literature. In section 3, we construct the GHCS and establish the resolution of the identity 
satisfied by these states by providing the appropriate weight function expressed by the Meijer G-functions \cite{mathai}, which solves the Stieltjes moment problem.  Section 4 deals with the analytical insights of the GHCSs through the reproducing kernel and the  analytic representation of a given function in the Hilbert space spanned by these states. In Section 5,    the expectation values of the observables describing the quantum system are derived. The
quantization of a complex plane,  known as  the Berezin-Klauder-Toeplitz quantization, (and also called  coherent state or anti-Wick quantization), using these states is investigated in Section 6. In Section 7,   the  
probability density and the time dependence of the GHCSs are discussed. 
%{\color{blue}
	Section 8 deals with the treament of thermodynamical
properties of the quantum system, with  the thermal expectations of the relevant observables determined from the Glauber-Sudarshan $P$-diagonal representation
of the density operator.
%}
%analyzed in the constructed GHCSs. 
\section{General method of GHCSs construction}
 We start with the following definition.
\bdfn
The  generalized associated  hypergeometric type CSs (GAH-CSs) are the CSs corresponding to  the $m^{\textrm{th}} $ derivative  $\Phi_{l,m}= \kappa^m \Phi_l^{(m)}$ of the classical orthogonal polynomials $\Phi_l$ satisfying the second order differential equation of hypergeometric type:
\beq \label{eq1}
\sigma(s)\Phi''_l(s)+\tau(s)\Phi'_l(s) + \lambda_l\Phi_l(s)=0,
\eeq
 where $\lambda_l = -{1 \over 2}l(l-1)\sigma'' - l\tau'$, $\kappa = \sqrt{\sigma}$,  with $\sigma$ a nonnegative function;  $\sigma$ and  $\tau$  are polynomials of at most second and exactly  first degrees, respectively.
\edfn
The $\Phi_{l,m}$, called associated hypergeometric-type functions (AHF), are solutions of the eigenvalue problem $H_m \Phi_{l,m} = \lambda_l \Phi_{l,m}$ where  the 
 Hamiltonian operator $H_m$ is expressed as a second order differential operator as follows:  
\beq\label{eq2}
H_m & = &-\sigma {d^2 \over ds^2} -\tau {d\over ds} + {m(m-2) \over 4}{{\sigma'}^2 \over \sigma} + {m\over 2}\tau{\sigma'\over \sigma}  - {1\over 2} m(m-2)\sigma'' - m \tau'.
\eeq
The $\Phi_{l,m}$ are orthogonal 
\beq \label{eq3} 
\int_a^b \Phi_{l,m}\Phi_{k,m}\, \rho\, d s
  =0,\qquad  l \ne k,\qquad l, \, k \in \{ m, m+1, m+2, \ldots \},
  \eeq
with respect to the positive weight function $\rho$ related to the  polynomial functions  $\sigma$ and $\tau$ by the Pearson's equation $(\sigma \rho)' = \tau
 \rho$, over  the interval $(a, b)$, which can be finite or infinite.
The  operator $H_m $   factorizes as:
   $$H_m - \lambda_m = A_m^\dag A_m,\quad
   H_{m+1}-\lambda_m = A_m A_m^\dag, $$
and     fulfills the intertwining relations
   $$H_m A_m^\dag = A_m^\dag H_{m+1}\quad \textrm{ and}\quad A_m H_m = H_{m+1}A_m. $$
The mutually formal  adjoint  first-order differential operators 
 $$A_m : {\cal H}_m \longrightarrow {\cal  H}_{m+1}\quad \textrm{ and}\quad 
A_m^\dag : {\cal H}_{m+1} \longrightarrow {\cal  H}_{m},$$ 
are    defined as \cite{cotfas}:
\beq
  A_m &= &\kappa(s){d \over d s} - m \kappa'(s)\quad \textrm{ and}\quad \\
   A_m^\dag &=& -\kappa(s)\frac{d}{d s} -   {\tau(s)\over \kappa(s)}   -(m-1)\kappa'(s),
   \eeq
with $\kappa = \sqrt{\sigma}$. 
 ${\cal H}_m$ is the Hilbert space of $\{\Phi_{k,m}\}_{k \ge m},$
  for  $m \in \N,$
  with respect to the inner product (\ref{eq3}).  We restrict ourselves to the case when   for each 
  $m\in \N$,  ${\cal H}_m$ is dense in the  Hilbert space ${\cal H} =\{ \varphi \in  L^2(\rho(s)ds)\}$ where $L^2$ 
  is the space of square integrable functions.
The following shape invariance relations are satisfied 
 \beq\label{eq4}
A_m A_m^\dag &=& A_{m+1}^\dag A_{m+1}+
r_{m+1},\quad \cr
r_{m+1} &=& \lambda_{m+1}-\lambda_m = -m\sigma'' -
\tau', 
\eeq
 where   eigenvalues $\lambda_l$ and eigenfunctions
$\Phi_{l,m}$ are:
 \beq \label{eq5}
 \lambda_l &=&\sum_{k=1}^l r_k,
\quad \cr
\Phi_{l,m}& =& \frac{A_m^\dag}{\lambda_l - \lambda_m}\frac{A_{m+1}^\dag}{\lambda_l - \lambda_{m+1}}\cdots
\frac{A_{l-2}^\dag}{\lambda_l - \lambda_{l-2}}\frac{A_{l-1}^\dag}{\lambda_l - \lambda_{l-1}}\Phi_{l,l}
\eeq
 for all $l\, \in \N$ and $ m\,  \in
\{ 0, 1, 2, \ldots, l-1\}$, $\Phi_{l,l}$ satisfying the
relation $\displaystyle A_l \Phi_{l,l} = 0.$\\
%lllllllllllllllllllllll
The annihilation  and  creation operators are defined as:
 \beq
 a_m, \, a_m^\dag: {\cal H}_m \longrightarrow
{\cal H}_m, \quad a_m = U_m^\dag A_m\  \textrm{ and}\  a_m^\dag = A_m^\dag U_m 
\eeq
within the unitary operator 
\beq
U_m: {\cal H}_m \longrightarrow {\cal H}_m,\quad U_m| l.
m \rangle = | l+1, m+1\rangle.
\eeq
The states $|l, m\rangle = \displaystyle {\Phi_{l,m}\over ||\Phi_{l,m}||}$ 
are defined for all $l \geq m$ and for each $m \in \N$.
 The mutually formal adjoint operators $a_m$ and $a_m^\dag$ 
act on the states $| l, m \rangle$ as 
\beq
   a_m | l, m \rangle &=&
\sqrt{\lambda_l - \lambda_m}\,| l-1, m \rangle\quad \textrm{
and}\cr
\quad a_m^\dag | l, m \rangle &=& \sqrt{\lambda_{l+1} -
\lambda_m}\,| l+1, m \rangle,\quad l \ge m,
 \eeq
and satisfy  the commutation relations:
{\small
\beq \label{eq6}
%\left\{ \ba{rcl}
[a_m, a_m^\dag] ={\cal R}_m ,  
[a_m^\dag, {\cal R}_m] = \sigma'' a_m^\dag,  
[a_m, {\cal R}_m] =  -\sigma'' a_m,
%\ea \right.
\eeq}
where ${\cal R}_m = -\sigma''N_m -\tau'$, $N_m: {\cal H}_m \longrightarrow {\cal
 H}_m$ is the number operator defined as $N_m \Phi_{l,m}= l
 \Phi_{l,m}.$
 Remark that, when $deg(\sigma) = 1$,  the algebra defined by the generators in (\ref{eq6}) is isomorphic to the Heisenberg-Weyl algebra \cite{cotfas}.
In addition to the commutation relations (\ref{eq6}),  we have
 \beq \label{eq7}
 A_m {\cal R}_m = {\cal R}_{m+1} A_m,
\eeq 
and the similarity transformation 
 \beq \label{eq8}
 U_m {\cal R}_m U_m^\dag =
 {\cal R}_{m+1}+\sigma'', \quad \textrm{for\,  all}\quad  m \, \in \N.
\eeq
Setting  for all $m\, \in\, \N$,
$| n \rangle = | m+n, m \rangle$,  $ e_n = \lambda_{m+n}-
\lambda_m,\quad m\, \in\, \N,$
 we obtain,  with $N_m = a_m^\dag a_m$:
 \beq  \label{eq9}
  a_m| n \rangle& = &\sqrt{e_n}| n -1 \rangle,\quad\cr
   a_m^\dag| n \rangle
&= &\sqrt{e_{n+1}}| n+1 \rangle, \quad\\
 (H_m - \lambda_m)| n \rangle &=&
e_n| n \rangle. \nonumber
\eeq 
The CSs for AHF were provided by Cotfas \cite{cotfas} as:
\beq \label{eq10}
|z\rangle =  {\cal N}(|z|^2) \sum_{n = 0}^\infty \frac{z^n}{\sqrt{\varepsilon_n}}| n\rangle, 
 \  {\cal N}(|z|^2) = \left[\sum_{n =0}^\infty \frac{|z|^{2n}}{\varepsilon_n}\right]^{-1/2}
 \eeq
 for any $z$ in the open disc ${\cal C}(O, {\cal R})$ with centre $O$ and
 radius 
\beq
 {\cal R} &= &\limsup_{n \to \infty }\sqrt[n]{\varepsilon_n} \ne 0
 \\
\quad \textrm{with} \quad  \varepsilon_n &= &\left\{ \ba{lcl l}  1 & & \textrm{if} & n = 0\\
 e_1 e_2 \cdots e_n & & \textrm{if} & n > 0 \ea \right..
 \eeq
   These CSs are eigenstates of the annihilation operator $a_m,$  i.e.,
$a_m | z\rangle = z|z\rangle$.\\
Introducing the right-inverse operators $A_m^{-1}$, $a_m^{-1}$, it was established in \cite{hounk-sodoga} that the CSs (\ref{eq10}) can be rewritten as 
\beq \label{eq11}
|z\rangle =  {\cal N}(|z|^2) \sum_{n = 0}^\infty {(z a_m^{-1})}^{n}| 0\rangle,
\eeq
with their generalization     given by
\beq \label{eq12}
 |z; {\cal R}_m\rangle & = &   \sum_{n = 0}^\infty {(z  {\cal R}_m  a_m^{-1})}^{n}| 0\rangle =
 \sum_{n = 0}^\infty \frac{z^n}{h_n({\cal R}_m)}| n\rangle
\eeq
where 
\beq \label{eq13}
 h_0({\cal R}_m) &= & 1, \\
 h_n({\cal R}_m) &=& \dis \frac{\sqrt{\varepsilon_n}}{\dis \prod_{k = 0}^{n-1}({\cal R}_m + k \sigma'')} \quad \textrm{for}\quad   n \ge 1.
\eeq
The states (\ref{eq12}) are  eigenstates
of $a_m$, 
\beq \label{eq14}
 a_m|z; {\cal R}_m \rangle = z
({\cal R}_m- \sigma'')|z; {\cal R}_m\rangle ,
 \eeq
 and  satisfy the second  order differential equation
{\small
 \beq \label{eq15}
\{a_m - z ({\cal R}_m - \sigma'') \}\frac{d}{d z}|z;
{\cal R}_m \rangle = ({\cal R}_m - \sigma'')|z; {\cal R}_m\rangle.
 \eeq}
Furthermore, we  generalized the CSs (\ref{eq12}) as: 
{\small
\beq \label{eq16}
 |z; {\cal R}_m\rangle &= &  \sum_{n = 0}^\infty {(z  f({\cal R}_m)  a_m^{-1})}^{n}| 0\rangle =   \sum_{n=0}^\infty \frac{z^n}{h_n({\cal
 R}_m)}|n\rangle 
 \eeq}
for any analytical function $f,$   where
{%\small
 \beq \label{eq17}
  h_0({\cal R}_m) &=& 1\cr 
 \  \textrm{and} \   h_n({\cal R}_m) &=& \frac{\sqrt{\varepsilon_n}}{\displaystyle
\prod_{k=0}^{n-1} f({\cal R}_m + k \sigma'')} \  \textrm{for}\    n \ge 1.
 \eeq}
The CSs  (\ref{eq16}) are  eigenstates
of $a_m$, 
 \beq \label{eq18}
 a_m|z; {\cal R}_m \rangle = z
f({\cal R}_m- \sigma'')|z; {\cal R}_m\rangle ,
 \eeq
 and  satisfy  the condition
 {\small
 \beq \label{eq19}
\{a_m - z f({\cal R}_m - \sigma'') \}{d\over dz}|z;
{\cal R}_m \rangle = f({\cal R}_m - \sigma'')|z; {\cal R}_m\rangle.
 \eeq}
 Taking into account the fact that ${\cal R}_m$ is an operator which acts on
 the states
 $|n \rangle$ as
 \beq \label{eq20}
 {\cal R}_m |n\rangle = [-(m+n)\sigma'' - \tau']|n
 \rangle = r_{m+n+1}|n\rangle ,
 \eeq
 we  rewrite the CSs (\ref{eq16}) under the form:
  \beq \label{eq21}
  |z; m\rangle = \sum_{n=0}^\infty \frac{z^n}{h_n(m)}|n\rangle, 
\eeq
\quad 
where 
\beq\label{eq21bis}
 h_0(m)& =& 1\cr
 \ \textrm{and} \  h_n(m) & = &
{\sqrt{\varepsilon_n} \over \dis \prod_{k=0}^{n-1} f(r_{m+n+1-k})} \quad \textrm{for} \quad n\ge 1.
\eeq
The properties (\ref{eq18}) and (\ref{eq19}) become
\beq\label{eq22}
a_m|z; m \rangle &=& z f(r_{m+n+2}'')|z; m\rangle ,\cr
\{a_m - z f(r_{m+n+2}) \}{d\over dz}|z;m \rangle &=& f(r_{m+n+2})|z; m\rangle,
 \eeq
respectively.
 We established in \cite{hounk-sodoga} that  the generalized coherent states (\ref{eq21}) verify the   properties of  label  continuity, 
overcompleteness,   temporal stability and   action identity.
\section{Generalized  hypergeometric coherent states  for associated
Jacobi polynomials and hypergeometric functions}

\subsection{The construction}
In  this section, we construct CSs for Jacobi associated fuctions and hypergeometric functions. The resolution of the identity satisfied by these states is discussed through the Stieltjes moment problem and solved using the Mellin transform. The appropriate solution is given  in terms of Meijer G functions. 
  
 The polynomial functions 
$\sigma(x)$ and $\tau(x)$ 
%delivered in the following cases as:
%  \begin{enumerate}
%  	\item {\bf 
  	corresponding to	Jacobi associated functions polynomials
  	%:}
  are
  	\begin{eqnarray}
  	\sigma(x) = 1-x^2, \qquad
    \tau(x)=(\zeta-\gamma)-(\gamma+\zeta+2)x
  	\end{eqnarray}
  %	\item {\bf
  	while for the
  		 hypergeometric functions polynomials
  	 %}
   they are given by
  	\begin{eqnarray}
  		\sigma(x) = (1-x)x, \quad
  	\tau(x) = (\zeta+1)-(\gamma+\zeta+2)x.
  	\end{eqnarray}
%  \end{enumerate}
 The derivatives of first order of $\tau(x)$ and second order of $\sigma(x)$ are provided as:
  	\begin{equation}\label{houn-sodo47}
  	\tau'(x)=-(\gamma+\zeta+2), \qquad	\sigma''(x)=-2, 
  	\end{equation}
respectively. For commodity, set
\begin{equation}\label{houn-sodo48}
\mu=\gamma+\zeta+2.
\end{equation}
The eigenvalues of the Hamiltonian  $H_m$, given in (\ref{eq2}),  with associated eigenvectors $\left\{\Phi_{l,m}\right\}_{l\ge 0},$ 
% engendrant l'espace de Hilbert $\mathcal{H}_m$ sont données à partir des relations
are obtained from (\ref{houn-sodo47}) and  (\ref{houn-sodo48}) as follows:
  \begin{eqnarray}\label{houn-sodo0056}
\lambda_{l}= -\frac{1}{2}l(l-1)\sigma''-l\tau'= l(l+\mu-1)
\end{eqnarray}
such that we get \cite{epjdhyper}:
\begin{equation}\label{eigval000}
e_n = 
 \lambda_{m+n} - \lambda_{m}  =  n(2m+n+\mu-1).
\end{equation}
%  La détermination des $\varepsilon_n$ conduit à 
% \begin{eqnarray}\label{houn-sodo51}
% \varepsilon_n =e_1\times e_2\times e_3\times \dots \times e_n

% \cr
% &=&1.(2m+\mu)2.(2m+\mu+1)3.(2m+\mu+2)\dots n.(2m+\mu+n-1)\cr
% &=&1.2.3\dots n.(2m+\mu)(2m+\nu+1)(2m+\mu+2)\dots (2m+\mu+n-1)\cr
% =n!\frac{\Gamma(2m+\mu+n)}{\Gamma(2m+\mu)}.
% \end{eqnarray}
% En remplaçant $\sigma''$ dans l'expression (\ref{houn-sodo32}), nous avons
%\begin{eqnarray}
%f(R_m+k\sigma'')&=&f(R_m-2k)
%\end{eqnarray}
%Supposons la fonction $f(R_m)=c$ avec c une constante nous constatons que
% \begin{eqnarray}
% \prod_{k=0}^{n-1}f(R_m-2k)&=&c^n
% \end{eqnarray}
% where $c$ is a  constant.
  The quantity $h_n(R_m)$ yields
\begin{equation}
h_n(R_m)=\sqrt{\Gamma(n+1)\frac{\Gamma(2m+n+\mu)}{c^{2n}\,\Gamma(2m+\mu)}}.
\end{equation}
In order to obtain $h_n(R_m)$ for any value of $c$, set
\begin{equation}
f(R_m)=f(R_m+\sigma'')=\sqrt{\left(-\frac{1}{2}R_m\right)\left(-\frac{1}{2}R_m\right)}
\end{equation}
with
\begin{equation}
\qquad R_m=-(m+n)\sigma''-\tau'=2\left(m+n+\frac{\mu}{2}\right).\nonumber
\end{equation}
Then, we obtain the product
 {\small
\begin{eqnarray}\label{houn-sodo57}
% \prod_{k=0}^{n-1}f(R_m+k\sigma'')&=&\prod_{k=0}^{n-1}f(R_m-2k)\cr
%&=&\prod_{k=0}^{n-1}\sqrt{\left[-\frac{R_m-2k}{2}\right]\left[-\frac{R_m-2k}{2}\right]}\cr
% &=&\prod_{k=0}^{n-1}\sqrt{\left[k-\frac{R_m}{2}\right]\left[k-\frac{R_m}{2}\right]}\cr
% &=&\sqrt{\left(-\frac{R_m}{2}\right)^2\left(1-\frac{R_m}{2}\right)^2\left(2-\frac{R_m}{2}\right)^2\dots\left(n-1-\frac{R_m}{2}\right)^2}\cr
\prod_{k=0}^{n-1}f(R_m+k\sigma'')&=&\sqrt{\frac{\Gamma\left(n-\frac{R_m}{2}\right)\Gamma\left(n-\frac{R_m}{2}\right)}{\Gamma\left(-\frac{R_m}{2}\right)\Gamma\left(-\frac{R_m}{2}\right)}}=c^n.
\end{eqnarray}
}
Fixing $\nu=\frac{\mu}{2},$ 
% et en mettant les relations (\ref{houn-sodo51}) et
 from (\ref{houn-sodo57})
% dans la relation (\ref{houn-sodo32}), nous avons
it comes :
\begin{eqnarray}
h_n(R_m)&=&\left[\Gamma(n+1)\frac{\Gamma(n+2m+2\nu)}{\Gamma(2m+2\nu)}\frac{\Gamma\left(-\frac{R_m}{2}\right)}{\Gamma\left(n-\frac{R_m}{2}\right)}\frac{\Gamma\left(-\frac{R_m}{2}\right)}{\Gamma\left(n-\frac{R_m}{2}\right)}\right]^{-\frac{1}{2}}.
\end{eqnarray}
% De plus, en remplaçant $R_m$ par $m$, nous obtenons
Thereby
% \begin{equation*}
% h_n(m)=\left[\Gamma(n+1)\frac{\Gamma(n+2m+\mu)}{\Gamma(2m+\mu)}\frac{\Gamma\left(-(m+n)-\frac{\mu}{2}\right)\Gamma\left(-(m+n)-\frac{\mu}{2}\right)}{\Gamma\left(n-(m+n)-\frac{\mu}{2}\right)\Gamma\left(n-(m+n)-\frac{\mu}{2}\right)}\right]^{-\frac{1}{2}},
% \end{equation*}
\begin{equation}\label{houn-sodo40}
h_n(m)=\left[\Gamma(n+1)\frac{(2m+2\nu)_n}{[\left(-(m+n)-2\nu\right)_n]^2}\right]^{-\frac{1}{2}}.
\end{equation}
% Nous constatons  que le coefficient  $h_n(m)$ est de type 
% \begin{equation}
% \sqrt{\frac{(b)_n}{(a_1)_n(a_2)_n}},
% \end{equation} où $a_1=a_2<0$ et $a_1+a_2-b=-2(m+n)+\mu-\mu=-2(m+n)$.
% \begin{pro}
% 	En partant du formalisme de construction (voir Chapitre 2) établissant que les états cohérents thermiques hypergéométriques généralisés sont définis sur un cercle de rayon $|z|$, alors la fonction de normalisation peut être exprimée en terme de fonctions  Meijer-$G$ telle que:
% 	\begin{equation}\label{houn-sodo0041}
% 	N(|z|^{2},m)=\frac{\Gamma(2m+\mu)}{[\Gamma(-m-n-\nu)]^2}\MeijerG{1,2}{2,2}{1+m+n+\nu,\,\,\,\,\,1+m+n+\nu}{0,\,\,\,\,\,\,\,\,\,\,\,\,\,\,\,1-2m-\mu}{-|z|^2}.
% 	\end{equation}
% \end{pro}
% {\bf Preuve}
% \begin{eqnarray}
% N(|z|^2,m)&=&\sum_{n=0}^{\infty}\frac{|z|^{2n}}{|h_n(m)|2}\cr
% &=&\sum_{n=0}^{\infty}\frac{|z|^{2n}}{\left[\Gamma(n+1)\frac{(2m+\mu)_n}{[\left(-(m+n)-\nu\right)_n]^2}\right]}\cr
% &=&\sum_{n=0}^{\infty}\frac{[\left(-(m+n)-\nu\right)_n]^2}{(2m+\mu)_n}\frac{|z|^{2n}}{\Gamma(n+1)}\cr
% &=&_2F_1(-m-n-\nu;-m-n-\nu;2m+\mu;|z|^{2n})\cr
% N(|z|^{2},m)&=&\frac{\Gamma(2m+\mu)}{[\Gamma(-m-n-\nu)]^2}\MeijerG{1,2}{2,2}{1+m+n+\nu,\,\,\,\,\,1+m+n+\nu}{0,\,\,\,\,\,\,\,\,\,\,\,\,\,\,\,1-2m-\mu}{-|z|^2}\nonumber.
% \end{eqnarray}$\quad\hfill{\square}$
% }

% \newpage 

%%%%%%%%%%%%%%%%%%%%%%%%%%%%%%%%%%%%%%%%%%%%%%%%%%%%ù
Then, the  related CSs $\ket{z;m}$ for both associated Jacobi
polynomials and hypergeometric functions  are provided as follows: 
\begin{equation}\label{houn-sodo0042}
\ket{z;m} =\left[N(|z|^{2},m)\right]^{-\frac{1}{2}}\sum_{n=0}^{\infty}\frac{z^n}{\sqrt{\frac{\Gamma(n+1)(2m+2\nu)_n}{[(-m-n-\nu)_n]^2}}}\ket{n}
\end{equation}
 where the normalization function is given in terms of Meijer's G function as 
 \beq
 N(|z|^{2},m) = \frac{\Gamma(2m+2\nu)}{[\Gamma(-m-n-\nu)]^2}\MeijerG{1,2}{2,2}{1+m+n+\nu,\,\,\,\,\,1+m+n+\nu}{0,\,\,\,\,\,\,\,\,\,\,\,\,\,\,\,1-2m-2\nu}{-|z|^2}.
 \eeq
\brmk
\begin{enumerate}
\item 
The constructed GHCSs (\ref{houn-sodo0042}) coincide with the GHCSs introduced by Appl and Schiller (\ref{applschill}), and correspond to the GPAH-CSs for $c \neq 1$ and for the number of added
quanta (or photons) $p=0$ \cite{epjdhyper}.
\item
In the case $c=1$, the GHCSs $\ket{z;m}$
are delivered by 
\beq\label{houn-sodo0036}
\ket{z;m}=\frac{1}{\sqrt{_0F_1(2m+2\nu;|z|^2)}}\sum_{n=0}^{\infty}\frac{z^n}{\sqrt{\Gamma(n+1)(2m+2\nu)_n}}\ket{n}
\eeq
with the normalization constant given as follows:
{\small
\begin{eqnarray}\label{normalz000}
N(|z|^2;m)&=&\sum_{n=0}^{\infty}\frac{|z|^{2n}}{|h_n(m)|^2} =   _0F_1(2m+2\nu;|z|^2)\cr
&=&\Gamma(2m+2\nu)|z|^{1-2m-2\nu}I_{2m+2\nu-1}(2|z|).
\end{eqnarray}
}
The GHCSs (\ref{houn-sodo0036}) coincide with the GPAH-CSs obtained in \cite{hounk-sodoga} for 
the number of added quanta (or photons) $p=0$. 
\end{enumerate}
\ermk
Moreover, they  are not orthogonal to each other since
\beq\label{overlap000}
\braket{z',m|z,m}
&=& \frac{1}{\left[\MeijerG{1,2}{2,2}{1+m+n+\nu,\,\,\,\,\,1+m+n+\nu}{0,\,\,\,\,\,\,\,\,\,\,\,\,\,\,\,1-2m-2\nu}{-|z'|^2}\right]^{\frac{1}{2}}}\frac{\MeijerG{1,2}{2,2}{1+m+n+\nu,\,\,\,\,\,1+m+n+\nu}{0,\,\,\,\,\,\,\,\,\,\,\,\,\,\,\,1-2m-2\nu}{-z{\bar{z'}}}}{\left[\MeijerG{1,2}{2,2}{1+m+n+\nu,\,\,\,\,\,1+m+n+\nu}{0,\,\,\,\,\,\,\,\,\,\,\,\,\,\,\,1-2m-2\nu}{-|z|^2}\right]^{\frac{1}{2}}}.\nonumber\\
\eeq
\subsection{Continuity in the labeling}
The GHCSs (\ref{houn-sodo0036}) are continuous in labeling
$z$. Indeed, the transformation of CSs
parameter $z' \longrightarrow z$ leads to the transformation of GHCSs $|z';m\rangle \longrightarrow |z;m\rangle$:
{\small
\beq
\mbox{If}\quad |z-z'|\longrightarrow 0 \quad \mbox{then} \||z,m\rangle - |z',m\rangle \|^2 
 = 2\left[1-\mbox{Re}\left(\langle z',m|z,m\rangle\right)\right]\longrightarrow 0, 
\eeq
}
where (\ref{normalz000}) and (\ref{overlap000}) together have been used.
\subsection{Resolution of the unity operator}
Next, a fundamental property of any CSs is the resolution of
unity operator. We have the following proposition:
\bpro\label{prop01}
The GHCSs $\ket{z;m}$ satisfy the following resolution of the identity 
\begin{equation}\label{ident000}
\int_{\mathbb{C}}\frac{d^2z}{\pi}\ket{z;m}\bra{z;m}W(|z|^2,m)=\mathbb{I_{\mathfrak H}}
\end{equation}
with the Hilbert space $\mathfrak H = span\{\ket{n}\}_{n= 0}^{\infty}$ identical to the Fock basis, 
where the weight function $W(|z|^2,m)$,  obtained through the Mellin transform, is given by 
\begin{eqnarray}\label{weightfunc000}
 	W(|z|^2,m)&=&\frac{1}{[\Gamma(-m-\nu)\Gamma(-m-n-\nu)]^2}\MeijerG{1,2}{2,2}{1+m+n+\nu,\,\,\,\,\,1+m+n+\nu}{0,\quad 1-2m-2\nu}{-|z|^2}\cr
 	&&\times\MeijerG{2,0}{2,2}{m+\nu;\,\,\,m+\nu}{0;\,\,\,2m+2\nu-1}{|z|^2}.
 	\end{eqnarray}
\epro
{\bf Proof.} See the Appendix.
$\hfill{\square}$
\vspace{0.05cm}
\begin{center}
\begin{figure}[h]
\resizebox{0.34\textwidth}{!}{%
\includegraphics{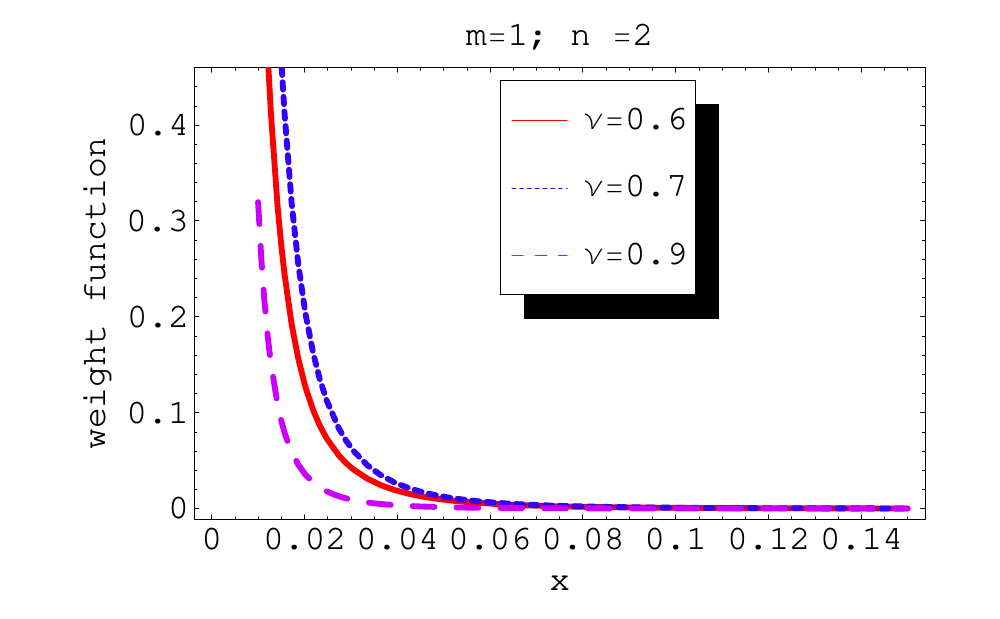}}
\resizebox{0.34\textwidth}{!}{%
\includegraphics{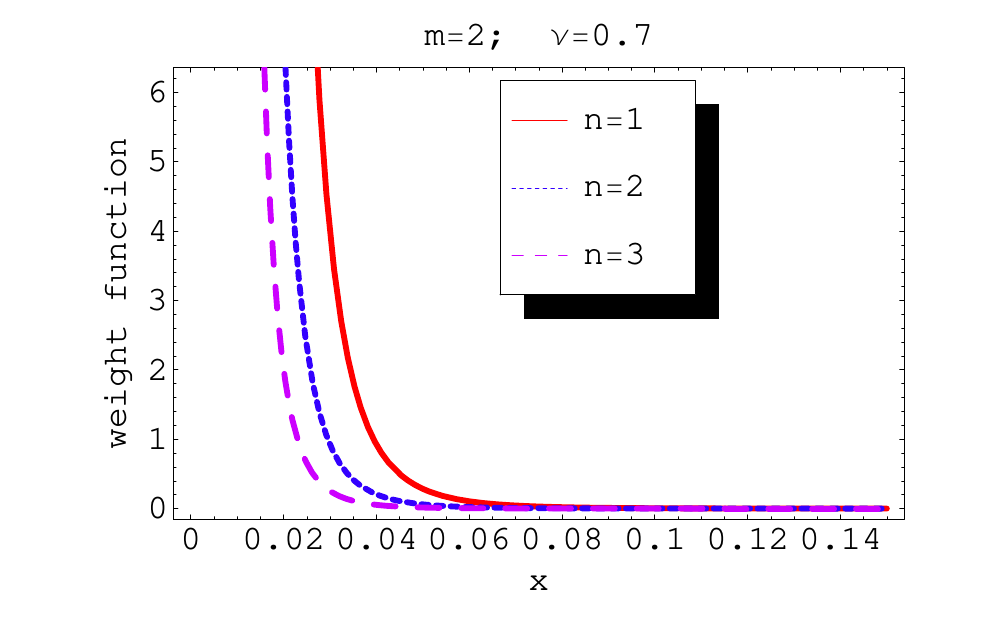}}\resizebox{0.34\textwidth}{!}{%
\includegraphics{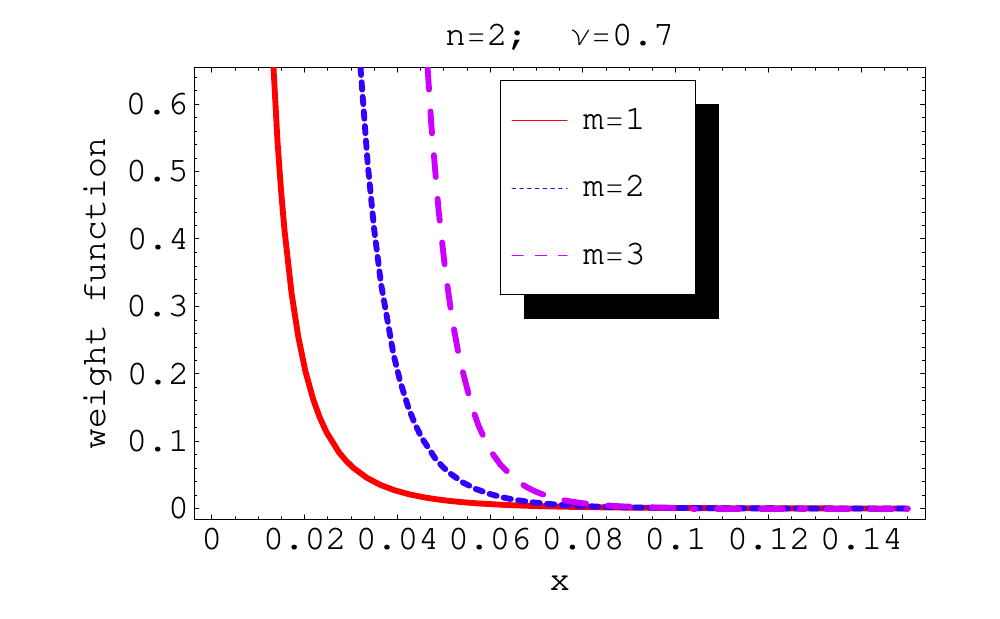}
}
\caption{\it \small %footnotesize 
%\color{blue}
 Plots of the  weight function (\ref{weightfunc000}) 
 versus  $x = |z|^2$:    with  the parameters $m = 1$, $n = 2$ 
 and for different  values of  $\nu$; %of the photon-added parameter $p$; 
   with  the parameters $m = 2$, $\nu=0.7$  and for different  values of  $n$;  with  the parameters $n = 2$, $\nu=0.7$  and for different  values of  $m$.
 } 
\label{weight-function}   
\end{figure} 
\end{center}
In Figure \ref{weight-function}, we plot the weight function (\ref{weightfunc000}) versus
$x = |z|^2$ for different values of $m, n$ and   $\nu$. All the curves
are positive, this confirms the positivity of the weight
function for the parameter $\nu > 0$. We also notice that the polynome parameter $\nu$ does not affect the general
behaviour of the curves but increases their asymptotic behaviour by taking smaller values.
\section{Reproducing kernel and analytic  representation}
\subsection{Reproducing kernel}
The overcompleteness of the constructed GHCSs $|z,m\rangle$ allows the study of their relation with the reproducing kernels \cite{ali-antoine-gazeau}. 

Define the quantity $K(z,z'):= \braket{z',m|z,m},$ 
%given, 
by using the connection between the  Meijer's G-functions and the modified Bessel functions of the first kind \cite{luke}-\cite{kampe},  as follows:
{\small
\beq\label{kern000}
K(z,z')&=&
\frac{\MeijerG{1,2}{2,2}{1+m+n+\nu,\,\,\,\,\,1+m+n+\nu}{0,\,\,\,\,\,\,\,\,\,\,\,\,\,\,\,1-2m-2\nu}{-z{\bar{z'}}}}
{\left[\MeijerG{1,2}{2,2}{1+m+n+\nu,\,\,\,\,\,1+m+n+\nu}{0,\,\,\,\,\,\,\,\,\,\,\,\,\,\,\,1-2m-2\nu}{-|z'|^2}\right]^{\frac{1}{2}}}
\frac{1}
{\left[\MeijerG{1,2}{2,2}{1+m+n+\nu,\,\,\,\,\,1+m+n+\nu}{0,\,\,\,\,\,\,\,\,\,\,\,\,\,\,\,1-2m-2\nu}{-|z|^2}\right]^{\frac{1}{2}}}\cr
&=&\left(\frac{|zz'|}{z\bar{z'}}\right)^{m+\nu-\frac{1}{2}}\frac{I_{2m+2\nu-1}(2\sqrt{z\bar{z'}})}{\sqrt{I_{2m+2\nu-1}(2|z|)I_{2m+2\nu-1}(2|z'|)}}.\nonumber \\
\eeq
}
$K(z,z')$ is a reproducing kernel. Indeed, we have the following result.
\bpro\label{prop03}
The following properties
\begin{enumerate}
\item [(i)] hermiticity $\overline{K(z,z')} = K(z',z),$ 
 \item  [(ii)] positivity $ K(z,z) > 0,$ and
 \item  [(iii)] idempotence  
 {\small$\displaystyle\int_{\mathbb{C}}K(z,z")K(z",z')\frac{W(|z"|^2)d^2z"}{\pi} = K(z,z')$}
\end{enumerate}
are satisfied by $K$ on the Hilbert spce $\mathfrak H$.
\epro
{\bf Proof.} 
See the Appendix.

$\hfill{\square}$
\subsection{Analytic representation in the GHCSs basis}
From the resolution of the identity property (\ref{ident000}), given $|\Psi\rangle \in \mathfrak H, $ we have 
\beq
\ket{\Psi}= \int_{\mathbb{C}}\frac{d^2z}{\pi}W(|z|^2,m)\Psi(z)\ket{z;m}
\eeq
where $\Psi(z):= \braket{z;m|\Psi}.$ Then, the  following reproducing property
\beq
\Psi(z) = \int_{\mathbb{C}}\frac{d^2z'}{\pi}W(|z'|^2,m)\Psi(z')K(z',z)
\eeq
is also satisfied. The Hilbert space $\mathfrak{H}$ can be represented as the Hilbert space of analytic functions in the variable $z$. 

Given a normalized state $\displaystyle\ket{\Phi}=\sum_{k=0}^{\infty}C_k\ket{k}, C_k \in \mathbb{C}$ on $\mathfrak{H}$, we obtain
\begin{eqnarray}
\braket{\bar{z};m|\Phi}
&=&\left[\frac{\Gamma(2m+2\nu)}{[\Gamma(-m-n-\nu)]^2}\MeijerG{1,2}{2,2}{1+m+n+\nu,\,\,\,\,\,1+m+n+\nu}{0,\,\,\,\,\,\,\,\,\,\,\,\,\,\,\,1-2m-2\nu}{-|z|^2}\right]^{-\frac{1}{2}}\cr
&&\times \sum_{n=0}^{\infty}C_n\frac{(-m-n-\nu)_n{z}^{n}}{\sqrt{\Gamma(n+1)(2m+2\nu)_n}}
\end{eqnarray}
such that the entire functions
\begin{eqnarray}
f(z,m)&=&\left[\frac{\Gamma(2m+2\nu)}{[\Gamma(-m-n-\nu)]^2}\MeijerG{1,2}{2,2}{1+m+n+\nu,\,\,\,\,\,1+m+n+\nu}{0,\,\,\,\,\,\,\,\,\,\,\,\,\,\,\,1-2m-2\nu}{-|z|^2}\right]^{\frac{1}{2}}\braket{\bar{z};m|\Phi}\cr
&=&\sum_{n=0}^{\infty}C_n\frac{(-m-n-\nu)_n{z}^{n}}{\sqrt{\Gamma(n+1)(2m+2\nu)_n}}
\end{eqnarray}
are analytic over the whole $z$ plane.   Then, from the resolution of the identity (\ref{ident000}), we can write
\beq
|\Phi \rangle = \int_{\C}\frac{d^{2}}{\pi}W(|z|^2,m)[N(|z|^2,m)]^{-1}f(\bar{z}, m)|z;m\rangle
\eeq
and express the scalar product of two states $|\Phi_1\rangle$ and $|\Phi_2\rangle$ given on $\mathfrak{H}$ by the formula 
% {\color{red}
{\small
\begin{equation}
\braket{\Phi_1|\Phi_2}=\int_{\mathbb{C}}\frac{d^2z}{\pi}W(|z|^2,m)[N(|z|^2,m)]^{-1}\overline{f_1(\bar z,m)}f_2(\bar{z},m)
\end{equation}
}
where 
\begin{eqnarray}
\overline{f_1(\bar z,m)}&=&\sum_{n=0}^{\infty}\overline{C_n}\frac{(-m-n-\nu)_n}{\sqrt{\Gamma(n+1)(2m+2\nu)_n}}z^n, \cr
f_2(\bar{z},m)&=&\sum_{k=0}^{\infty}C_k\frac{(-m-k-\nu)_k}{\sqrt{\Gamma(k+1)(2m+2\nu)_k}}\bar{z}^{k}.
\end{eqnarray}
% }
\section{Expectation values}
The constructed  CSs can be used in different physical applications to calculate
the expectation (mean) values of any significant physical observable $\mathcal O$ which
characterizes the quantum system embedded in the considered potential. We have the following statement.
%{\color{blue}
\bpro \label{prop02}
From  the definition of the CSs in (\ref{houn-sodo0042}), the mean value of a physical observable $\mathcal O$ in the  generalized CSs for hypergeometric type functions  $\ket{z;m}$  is obtained as
	\begin{eqnarray}
	\braket{{A}}_{z,m}&=&\left[\frac{\Gamma(2m+2\nu)}{[\Gamma(-m-n-\nu)]^2}\MeijerG{1,2}{2,2}{1+m+n+\nu,\,\,\,\,\,1+m+n+\nu}{0,\,\,\,\,\,\,\,\,\,\,\,\,\,\,\,1-2m-2\nu}{-|z|^2}\right]^{-1}\cr
	&&\times
	\sum_{n=0}^{\infty}\frac{z^{2n}\bra{n}A\ket{n}} {\frac{\Gamma(n+1)(2m+2\nu)_n}{[(-m-n-\nu)_n]^2}}.
	\end{eqnarray}
Then, we get 
\begin{eqnarray}
\braket{N}_{z,m}
=-C_1^{(n)}\frac{\MeijerG{1,2}{2,2}{1+m+n+\nu,\,\,\,\,1+m+n+\nu}{1,\,\,\,\,\,\,\,\,1-2m-2\nu}{-|z|^2}}{\MeijerG{1,2}{2,2}{1+m+n+\nu,\,\,\,\,1+m+n+\nu}{0,\,\,\,\,\,\,\,\,1-2m-2\nu}{-|z|^2}},
\end{eqnarray}
and 
\begin{eqnarray}
\braket{{N}^{2}}_{z,m}
 &=&\frac{1}{\MeijerG{1,2}{2,2}{1+m+n+\nu,\,\,\,\,1+m+n+\nu}{0,\,\,\,\,\,\,\,\,1-2m-2\nu}{-|z|^2}}\left[-C_1^{(n)}\MeijerG{1,2}{2,2}{1+m+n+\nu,\,\,\,\,1+m+n+\nu}{1,\,\,\,\,\,\,\,\,1-2m-2\nu}{-|z|^2} \right.\cr && \left. + C_2^{(n)}\MeijerG{1,2}{2,2}{1+m+n+\nu,\,\,\,\,1+m+n+\nu}{2,\,\,\,\,\,\,\,\,1-2m-2\nu}{-|z|^2}\right],
\end{eqnarray}
providing the intensity correlation 
\begin{eqnarray}
(g^{(2)})_{z,m}=\MeijerG{1,2}{2,2}{1+m+n+\nu,\,\,\,\,1+m+n+\nu}{2,\,\,\,\,\,\,\,\,1-2m-2\nu}{-|z|^2}\frac{\MeijerG{1,2}{2,2}{1+m+n+\nu,\,\,\,\,1+m+n+\nu}{0,\,\,\,\,\,\,\,\,1-2m-2\nu}{-|z|^2}}{\MeijerG{1,2}{2,2}{1+m+n+\nu,\,\,\,\,1+m+n+\nu}{1,\,\,\,\,\,\,\,\,1-2m-2\nu}{-|z|^2}}
\end{eqnarray}
with $C_1^{(n)} = C_2^{(n)} = 1$.
\epro
%}
{\bf Proof.} See  the Appendix.

$\hfill{\square}$\section{Quantization with the generalized hypergeometric coherent states}
In this paragraph, we deal  with the general procedure described in \cite{gazeaubook} and used, for example, in our previous  works 
\cite{aremua-gazeau-hk}-\cite{aremuaromp}. For more details, one may also consult references quoted therein.  
\subsection{Coherent State Quantization: General Scheme}
Let $X$ be a set of parameters equipped with a measure $\mu$ and let $L^2(X, \mu)$ be its associated Hilbert space  of 
complex-valued square integrable functions with respect to $\mu$. Let us choose 
in $L^2(X, \mu)$ a finite or countable  orthonormal set $\mathcal{O}=\{\phi_n\, , \, n =0,1,2,\dots \}$,
% with $\mathcal{F}$ 
% some countable set ($\sim \N$ or $\sim \Z$ ...),
\beq\label{eqI1}
\langle  \phi_m | \phi_n \rangle  = \int_{X}\overline{\phi_m(x)}\, \phi_n(x)\, \mu(dx) = \delta_{mn}\, ,
\eeq
 obeying the (crucial) 
 condition:
\beq\label{eqI2}
0< \sum_{n} | \phi_n (x)|^2 := \mathcal{N}(x) < \infty \,  \quad \mbox{a.e.}\, . 
\eeq
Let $\mathcal{H}:=\overline{\mathrm{span}(\mathcal{O})}$ in the Hilbert space $L^2(X, \mu)$  be a separable 
complex Hilbert space with orthonormal basis $\{|e_n\rangle\, , n =0,1,2,\dots \}$, 
in one-to-one 
correspondence with the elements of  $\mathcal{O}=\{\phi_n\, , \, n =0,1,2,\dots \}$.  
% In particular,  it can be chosen as  the Hilbert subspace 
% {\color{blue}$\mathcal{K}_{\mathcal{O}}\deq \overline{\mathrm{span}(\mathcal{O})}$ in $L^2(X, \mu)$} itself. 
One defines  the  family of  states $\mathcal{F}_{\mathcal{H}}= \{|x\rangle\, , \, x \in X \}$ in $\mathcal{H}$ as:
\beq\label{eqI3}
|x\rangle = \frac{1}{\sqrt{\mathcal{N}(x)}}\sum_n \overline{\phi_n(x)}\, |e_n\rangle\ \in \mathcal  H.
\eeq
From conditions (\ref{eqI1}) and (\ref{eqI2}) these CS are normalized, $\langle  x| x\rangle  = 1$ 
and resolve the identity in 
$\mathcal{H}$:
\beq\label{eq14}
\int_X  \,\mathcal{N}(x) \, |x\rangle \langle  x |\; \mu(dx) = \mathbb I_{{\mathcal H}}\, .
\eeq
The  relation (\ref{eq14}) allows us to implement a \emph{coherent state   quantization} of the set of parameters $X$ 
by associating to a function $X \ni x \mapsto f(x)$ that satisfies appropriate conditions 
  the   operator $ A_f$
in $\mathcal{H}$ as:
\begin{equation}\label{eqI5}
f(x) \mapsto A_f  := \int_X\,\mathcal{N}(x) \, f(x)\, |x\rangle \langle x |\; \mu(dx).
\end{equation}
The matrix elements of $A_f$ with respect to the basis $|e_n\rangle$ are given by
\begin{eqnarray}
\label{matelgen}
\left( A_f\right)_{nm} &=& \langle e_n| A_f|e_{m}\rangle \cr
&=& \int_X \, f(x) \, \overline{\phi_n(x)}\, \phi_{m}(x)\;  \mu(dx). 
\end{eqnarray}
The operator $A_f$ is 
 \begin{enumerate}
  \item  symmetric if $f(x)$ is real valued, 
  \item bounded  if $f(x)$ is bounded, and
  \item self-adjoint  if $f(x)$ is real semi-bounded (through \\ Friedrich's 
  extension or self adjoint extension). 
 \end{enumerate}
\subsection{Quantization of elementary classical observables}
The resolution of the identity provided by  (\ref{ident000}) allows us to implement the CS quantization (also named Berezin-Klauder-Toeplitz
or anti-Wick quantization) of the complex plane by associating a function $\mathbb{C}
\ni z\mapsto f(z)$. For this purpose, let us define the operator on the Hilbert space $\mathfrak{H}$
\beq
f(z) \mapsto A_f = \int_{\mathbb{C}}f(z)\ket{z;m}\bra{z,m}\frac{d^2z}{\pi}W(|z|^2,m)
\eeq
such that 
\beq\label{csquant000}
A_f &:=& \sum_{n=0}^{\infty}\sum_{k=0}^{\infty}\frac{\ket{n}\bra{k}}{\sqrt{\frac{\Gamma(n+1)(2m+2\nu)_n}{[(-N_m-\nu)_n]^2}\frac{\Gamma(k+1)(2m+2\nu)_k}{[(-N_m^{\dagger}-\nu)_k]^2}}}\int_{\mathbb{C}} [N(|z|^2,m)]^{-1} f(z) z^{n}\bar{z}^{k}\frac{d^2z}{\pi}W(|z|^2,m)
\eeq
where the operators $N_m, N_m^{\dagger}$ act on the Fock Hilbert space $\{|n\rangle\}^{\infty}_{n=0}$ as delivered in (\ref{eq9}).

The Berezin-Klauder-Toeplitz quantization of the elementary classical variables $z$ and
$\bar z$ is realized via the maps $z \mapsto A_z$  and $\bar z \mapsto A_{\bar z}$  defined on the Hilbert $\mathfrak{H}$. Then,  after some algebra, we obtain in the complex plane 
\beq
A_z &=& \sum_{n=0}^{\infty}(-m-n-1-\nu)\sqrt{(n+1)(2m+2\nu+n)}\ket{n}\bra{n+1},
\eeq
\beq
A_{\bar{z}} &=&\sum_{n=0}^{\infty}(-m-n-\nu)\sqrt{n(2m+2\nu+n-1)}\ket{n}\bra{n-1},
\eeq
where the matrix elements 
\beq
(A_z)_{k,n} &=& \sum_{n=0}^{\infty}\sum_{k=0}^{\infty}\int_{0}^{2\pi}\int_{0}^{\infty}\frac{rdrd\theta}{\pi}[N(r^2,m)]^{-1}\frac{e^{i(n+1-k)\theta}r^{n+1+k}W(r^2,m)}{\sqrt{\frac{\Gamma(n+1)(2m+\mu)_n}{[(-N_m-\nu)_n]^2}\frac{\Gamma(k+1)(2m+\mu)_k}{[(-N_m^{\dagger}-\nu)_k]^2}}}\ket{n}\bra{k},
\eeq
\beq
(A_{\bar{z}})_{k,n} &=&\sum_{n=0}^{\infty}\sum_{k=0}^{\infty}\int_{0}^{2\pi}\int_{0}^{\infty}\frac{rdrd\theta}{\pi}[N(r^2,m)]^{-1}\frac{e^{i(n-k-1)\theta}r^{n+1+k}W(r^2,m)}{\sqrt{\frac{\Gamma(n+1)(2m+\mu)_n}{[(-m-n-\nu)_n]^2}\frac{\Gamma(k+1)(2m+\mu)_k}{[(-m-k-\nu)_k]^2}}}\ket{n}\bra{k},
\eeq
are together obtained   from (\ref{csquant000}), and the following  relations 
\beq
&&z =re^{i\theta}, \qquad\bar{z}=re^{-i\theta}\qquad\text{and}\qquad\frac{d^2z}{\pi}=\frac{rdrd\theta}{\pi}\cr
&&\int_{0}^{2\pi}e^{i(n-m)\theta}
d\theta = 
\left\{
              \begin{array}{lll}
              0 \quad \mbox{if}  \quad m \neq n, \\
               \\
              2\pi 
\quad \mbox{if}  \quad m = n
               \end{array}
\right.
% 	\nonumber\\
% 	|z|^2=x\Rightarrow r^{2n}=x^n\qquad\qquad
% 	dx=2|z|d|z|\Rightarrow \,\,dx=2rdr,\nonumber
\eeq
are used.

The  commutator of the operators $A_z$ and $A_{\bar{z}}$ takes the form
\beq
[A_z,A_{\bar{z}}]&=&
\sum_{n=0}^{\infty}\left\{(-m-n-1-\nu)^2(n+1)(2m+2\nu+n)  -(-m-n-\nu)^2 n(2m+2\nu+n-1)\right\}\cr
&&\times\ket{n}\bra{n}.
\eeq
For the quantization of $f(z) = |z|^2$ , using the integral formula (\ref{csquant000}), we obtain the corresponding operator
\beq
A_{|z|^2}&=&\sum_{n=0}^{\infty}\sum_{k=0}^{\infty}\int_{0}^{2\pi}\int_{0}^{\infty}\frac{rdrd\theta}{\pi}[N(r^2,m)]^{-1}\frac{e^{i(n-k-1)\theta}r^{n+1+k}}{\sqrt{\frac{\Gamma(n+1)(2m+\mu)_n}{[(-m-n-\nu)_n]^2}\frac{\Gamma(k+1)(2m+\mu)_k}{[(-m-k-\nu)_k]^2}}}W(r^2,m)\ket{n}\bra{k}\cr
&=&\sum_{n=0}^{\infty}(n+1)(2m+\mu+n)\ket{n}\bra{n}.
\eeq
Its commutators with $A_z$ and $A_{\bar z}$ yield
\beq
[A_z,A_{|z|^2}]&=& -2\sum_{n=0}^{\infty}(m+\nu+n+1)^2\sqrt{(n+1)(2m+2\nu+n)}\ket{n}\bra{n+1},
\eeq
\beq
[A_{\bar{z}},A_{|z|^2}]&=& 2 \sum_{n=0}^{\infty}(m+\nu+n)^2\sqrt{n(2m+2\nu+n-1)}\ket{n}\bra{n-1},
\eeq
respectively.
\section{Probability density and time evolution}
This section deals with  the semi-classical character of the  GHCSs. We
analyse how these states do evolve in time under the action of the time evolution
operator provided by the physical Hamiltonian describing the quantum system.

We start with the overlap (\ref{overlap000})  expressed in terms of the  modified Bessel functions of the first kind as follows:
\begin{eqnarray}
\braket{z',m|z,m}
&=&\left(\frac{|zz'|}{z\bar{z'}}\right)^{m+\nu-\frac{1}{2}}\frac{I_{2m+2\nu-1}(2\sqrt{z\bar{z'}})}{\sqrt{I_{2m+2\nu-1}(2|z|)I_{2m+2\nu-1}(2|z'|)}}.
\end{eqnarray}
Then, taking a normalized state $|z_{0};m\rangle$, the related phase space distribution is provided through  the
probability density: 
\beq\label{evolmap}
z\rightarrow\varrho_{z_0}(z)&:=& |\langle z;m|z_0;m\rangle|^2 \cr
&=& \frac{I_{2m+2\nu-1}(2\sqrt{z\bar{z'}})I_{2m+2\nu-1}(2\sqrt{\bar{z}z'})}{I_{2m+2\nu-1}(2|z|)I_{2m+2\nu-1}(2|z'|)}.
\eeq
\brmk
The expression (\ref{evolmap})  is analogue to the probability density  defined for BGCSs 
\beq{\label{gen05}}
|z\rangle_{m} = \frac{|z|^{m/2}}{\sqrt{I_{m}(2|z|)}}
\sum_{n=m}^{+\infty}\frac{z^{n-m}}{\sqrt{\Gamma(n-m+1)\Gamma(n+1)}}|n,m\rangle
\eeq
built in \cite{aremuaromp},  given as follows:
\beq
\overbrace{\varrho_{z_0}}(z) :=  |_{m}\langle z|z_0\rangle_{m}|^2
&=& \frac{I_m(2\sqrt{z_0 \bar z}) I_m(2\sqrt{\bar z_0  z})}{I_{m}(2|z|)I_{m}(2|z_0 |)}
\eeq
also provided through  modified Bessel functions of the first kind. Thus,   we can emphasize that the GHCSs (\ref{houn-sodo0036})  show similar time evolution behaviour as the BGCSs (\ref{gen05}).
\ermk
Then, the associated time evolution behaviour is supplied by 
\beq\label{tempstab}
z\rightarrow\varrho_{z_0}(z,t)=|\bra{z;m}e^{-iH_mt}\ket{z_0;m}|^2.
\eeq
After acting the evolution operator $U(t) = e^{-iH_mt}$,  with $H_m$ provided by (\ref{eq2}) and its eigenvalues  $e_n = n(n +2\nu +2m - 1)$ with $\mu = 2\nu$ (see (\ref{eigval000})),  on the GHCSs $|z_0;m\rangle$, we obtain
\begin{eqnarray}\label{temp000}
|z_0;t;m\rangle &=& e^{-iH_mt}\ket{z_0;m}\cr
&=&[N(|z_0|^2,m)]^{-\frac{1}{2}}\sum_{n=0}^{\infty}\frac{(-m-n-\nu)_n(z_0)^n}{\sqrt{\Gamma(n+1)(2m+2\nu)_n}}e^{-iH_mt}\ket{n}\cr
&=&[N(|z_0(t)|^2,m)]^{-\frac{1}{2}}\sum_{n=0}^{\infty}\frac{(-m-n-\nu)_n(z_0 e^{-i(n +2\nu +2m - 1)t})^n}{\sqrt{\Gamma(n+1)(2m+2\nu)_n}}\ket{n}.
\end{eqnarray}
Then, in the basis $\{|e^{-in^2t}n\rangle:= |\Phi_n(t)\rangle = |e^{-i \theta_n(t)}n\rangle,\\ 
\theta_n(t) = n^2t\}_{n=0}^{\infty}$, the equation (\ref{temp000}) becomes
\begin{eqnarray}
\overbrace{|z_0;t;m\rangle} &=:& e^{-iH_mt}\ket{z_0;m}\cr
&=& [N(|z_0(t)|^2,m)]^{-\frac{1}{2}}\sum_{n=0}^{\infty}\frac{(-m-n-\nu)_n(z_0(t))^n}{\sqrt{\Gamma(n+1)(2m+2\nu)_n}}|\Phi_n(t)\rangle\cr
&=& |z_0(t);m\rangle
\end{eqnarray}
where $z_0(t):=z_0 e^{-i(2\nu +2m - 1)t}$.
Then, by recasting  the GHCSs in the basis $\{|\Phi_n(t)\rangle\}_{n=0}^{\infty}$ as follows:
\begin{eqnarray}
\overbrace{|z;m\rangle }&=& [N(|z|^2,m)]^{-\frac{1}{2}}\sum_{n=0}^{\infty}\frac{(-m-n-\nu)_n z^n}{\sqrt{\Gamma(n+1)(2m+2\nu)_n}}|\Phi_n(t)\rangle
\end{eqnarray}
we get from (\ref{evolmap})
{\small
\beq
\varrho_{z_0}(z,t)&:=&|\overbrace{\bra{z;m}}\overbrace{z_0;t;m\rangle}|^2\cr 
&=& \frac{I_{2m+2\nu-1}(2\sqrt{\bar{z}z_0(t)})I_{2m+2\nu-1}(2\sqrt{z\overline{z_0(t)}})}{I_{2m+2\nu-1}(2|z|)I_{2m+2\nu-1}(2|z_0(t)|)}.
\eeq
}
Thereby, the time dependence of a given GHCSs $|z;m\rangle$ is realized as 
\beq\label{tempprop}
\overbrace{|z;t;m\rangle} &=& e^{-iH_mt}|z;m\rangle \cr
&=& |z(t);m\rangle, \; z(t):= e^{-i(2\nu +2m - 1)t} z.
\eeq
The relation (\ref{tempprop}) shows that the time evolution of the GHCSs $|z;m\rangle$ reduces to
a rotation in the complex plane given by $z \mapsto z(t)=  e^{-i(2\nu +2m - 1)t} z$.  Therefore, the semi-classical feature of the GHCSs is given by (\ref{tempstab}),
while the temporal stability property is highlighted by the relation (\ref{tempprop}). 
The latter
asserts that the temporal evolution of any GHCS always remains a GHCS, and fixes the
phase behaviour of the GHCSs $|z;m\rangle$  with the factor $ e^{-i(2\nu +2m - 1)t}$.
%{\color{blue}
\section{Thermal properties of the GHCSs}
This section furnishes a description of statistical properties of the GHCSs for the  model in the situation of a thermal equilibrium. Consider a quantum gas of the system in the thermodynamic
equilibrium with a reservoir at temperature $T$,  which satisfies a quantum canonical
distribution. The corresponding normalized density operator
is given by
{\small
\begin{eqnarray}
	\rho^{(m)}&=&\frac{1}{Z(\beta)}\sum_{n=0}^{\infty}e^{-\beta E_n}\ket{n}\bra{n}, \; E_n = n(n+\mu+1), 
	\end{eqnarray}
	}
	where the partition function denoted $Z(\beta)$ is taken as the normalization constant with its expression:
	\begin{equation}
    Z(\beta) = \sum_{n=0}^{\infty}e^{-\beta n(n+\mu+1)}
	\end{equation}
ensuring the normalization condition: $\mbox{Tr}(\rho^{(m)})=1$. 

%{\color{blue}  
The density operator GS-$P$-representation (also known as the diagonal expansion) is given by
\begin{equation}
\rho=\int_{\mathbb{C}}\frac{d^2z}{\pi}W(|z|^2,m)\ket{z;m}P(|z|^2,m)\bra{z;m},
\end{equation}
with the quasi-distribution function  $P(|z|^2,m)$ given, by the normalization condition 
\begin{equation}
\frac{1}{\pi}\int W(|z|^2,m)P(|z|^2,m)d^2z = 1,
\end{equation}
 as 
\begin{eqnarray}
P(|z|^2,m)&=& e^{\beta\mu}\sum_{k=0}^{\infty}\frac{\beta^k}{k!}\left(\frac{d}{d a}\right)^{2k}
\frac{\left[e^{a}\MeijerG{2,0}{2,2}{m+\nu,\,\,\,\,\,\,\,m+\nu}{0,\,\,\,\,\,\,\,\,\,\,\,2m+\mu-1}{e^{a}\,|z|^2}\right]}{\MeijerG{2,0}{2,2}{m+\nu,\,\,\,\,\,\,\,m+\nu}{0,\,\,\,\,\,\,\,\,\,\,\,2m+\mu-1}{|z|^2}},
\end{eqnarray}
where  $a = \beta(\mu +1)$.
%}

Using the identity (see \cite{popov1,popov3}) 
\begin{eqnarray}\label{houn-sodo0067}
	e^{-\beta E_n}
	&=&e^{-n\,\beta (\mu +1)}\,\sum_{k=0}^{\infty}\frac{\beta^k}{k!}(n)^{2k}
	\end{eqnarray} 
%{\color{red}and setting $a = \beta(\mu +1)$,}
 we obtain the thermal average of the integer powers of  number operator $\langle N^s \rangle$ in terms of Meijer's G-functions:
\begin{eqnarray}\label{houn-sodo0083}
	\braket{e^{\varepsilon N}}_{\ket{z;m}} = \frac{\MeijerG{1,2}{2,2}{1+m+n+\nu,\,\,\,\,\,1+m+n+\nu}{0,\,\,\,\,\,\,\,\,\,\,\,\,\,\,\,1-2m-\mu}{-e^{\varepsilon}|z|^2}}{\MeijerG{1,2}{2,2}{1+m+n+\nu,\,\,\,\,\,1+m+n+\nu}{0,\,\,\,\,\,\,\,\,\,\,\,\,\,\,\,1-2m-\mu}{-|z|^2}}.
	\end{eqnarray}
	Thereby
	\begin{eqnarray}\label{houn-sodo0084}
	\lim\limits_{\substack{\varepsilon\rightarrow 0}}\frac{\partial^{s}}{\partial \varepsilon^{s}}\braket{e^{\varepsilon {N}}}_{\ket{z;m}}&=&\braket{{N}^{s}}_{\ket{z;m}}\lim\limits_{\substack{\varepsilon\rightarrow 0}}\braket{e^{\varepsilon {N}}}\cr
	&=&\braket{{N}^{s}}_{\ket{z;m}}.
	\end{eqnarray}
Then, from
\begin{eqnarray}
	\braket{{N}^{s}}&=&\int_{\mathbb{C}}\frac{d^2z}{\pi}\,W(|z|^2,m)\,P(|z|^2,m)\,\braket{{N}^{s}}_{\ket{z;m}}\cr
	&=& \lim\limits_{\substack{\varepsilon\rightarrow 0}}\frac{\partial^{s}}{\partial \varepsilon^{s}}\int_{\mathbb{C}}\frac{d^2z}{\pi}\,W(|z|^2,m)\,P(|z|^2,m)\braket{e^{\varepsilon {N}}}_{\ket{z;m}},
	\end{eqnarray}
	we finally get 
	\begin{equation}
	\braket{{N}^{s}}
	= \sum_{n=0}^{\infty}e^{\beta\mu}\sum_{k=0}^{\infty}\frac{\beta^k}{k!}\left(\frac{d}{d a}\right)^{2k}\lim\limits_{\substack{\varepsilon\rightarrow 0}}\frac{\partial^{s}}{\partial \varepsilon^{s}}(e^{\varepsilon})^n\,(e^{-a})^{n}.
	\end{equation}	
	In this manner, the thermal average of the first two powers of the number operator are 
\begin{equation}
	\braket{N} = 1+2e^{\beta(2-\mu)},\qquad
	\braket{N^2}
	= 1+4e^{\beta(2-\mu)}.
	\end{equation}
	Therefore, the thermal second-order correlation function $g^{(2)}$ and the thermal Mandel parameter are obtained as follows:
	\begin{equation}
	g^{(2)}=\frac{\braket{N^2}-\braket{N}}{\braket{N^2}} = \frac{2e^{\beta(2-\mu)}}{(1+2e^{\beta(2-\mu)})^2}, 
	\end{equation}
	\begin{equation}
	Q = \braket{N}(g^{(2)}-1) = -\left[1+4\frac{e^{2\beta(2-\mu)})}{1+2e^{\beta(2-\mu)}}\right].
	\end{equation}
%}
\section{Concluding remarks}
In this work, we have first explored the building method, developed in \cite{cotfas}-\cite{epjdhyper},  for generalized associated hypergeometric coherent states, and, then, proposed a   construction method for generalized hypergeometric CSs (GHCSs)
for associated Jacobi
polynomials and hypergeometric functions.  The constructed GHCSs  coincide with the GHCSs introduced by Appl and Schiller, and correspond to the GPAH-CSs for $c \neq 1$ and for the number of added
quanta (or photons) $p=0$ \cite{epjdhyper}. The resolution of the identity property is established through a Stieltjes moment problem solved by an appropriate weight function,
in terms of product of Meijer's G functions, by using the Mellin transform. Then, the  analytical features  of the GHCSs are discussed through the reproducing kernel and the analytic representation of
a given function in the Hilbert space spanned by these CSs. Next, the expectation values of the observables describing the quantum model have been derived in the constructed GHCSs basis. Besides, the
CS quantization procedure, known as the Berezin-Klauder-Toeplitz quantization, (and also called coherent state, or
anti-Wick quantization),  has been applied in the complex plane by using the basis of the GHCSs. The study of the properties of the GHCSs has been also carried out by the analysis of their time dependence  under the action of the time evolution operator elaborated from the quantum Hamiltonian and a  probability density. 
%{\color{blue}
	Using the GS-$P$-representation also known as the  diagonal representation of the density operator,
the relevant thermodynamical properties of the quantum system  have been investigated and discussed
in the GHCSs basis.
%}
\section*{Appendix}

{\bf Proof of Proposition \ref{prop01}}

From the definition of the CSs (\ref{houn-sodo0042}), we have
\begin{eqnarray}\label{resolv001}
\int_{\mathbb{C}}\frac{d^2z}{\pi}\ket{z;m}\bra{z;m}W(|z|^2,m)
% 
% &=&\sum_{n=0}^{\infty}\sum_{k=0}^{\infty}2\delta_{n,k}\cr
% &&\times\frac{r^{n+k}}{\sqrt{\frac{\Gamma(n+1)(2m+2\nu)_n}{[(-m-n-\nu)_n]^2}\frac{\Gamma(p+1)(2m+2\nu)_k}{[(-m-k-\nu)_k]^2}}}rdr\cr
% &&\times W(r^2,m)\ket{n}\bra{k}
=\sum_{n=0}^{\infty}\int_{0}^{\infty}[N(x,m)]^{-1}\frac{x^{n}W(x,m)dx}{\frac{\Gamma(n+1)(2m+2\nu)_n}{[(-m-n-\nu)_n]^2}}\ket{n}\bra{n}
\end{eqnarray}
% Pour $n=k$, on a:
% \medskip
% \begin{equation}
% \int_{\mathbb{C}}\frac{d^2z}{\pi}\ket{z;m}\bra{z;m}W(|z|^2,m)
% &=&\sum_{n=0}^{\infty}\int_{0}^{\infty}[N(r^2,m)]^{-1}\frac{r^{2n}}{\frac{\Gamma(n+1)(2m+\mu)_n}{[(-m-n-\nu)_n]^2}}2rdr \,W(r^2,m)\ket{n}\bra{n}\cr
% \int_{\mathbb{C}}\frac{d^2z}{\pi}\ket{z;m}\bra{z;m}W(|z|^2,m)
% =\sum_{n=0}^{\infty}\int_{0}^{\infty}[N(x,m)]^{-1}\frac{x^{n}}{\frac{\Gamma(n+1)(2m+\mu)_n}{[(-m-n-\nu)_n]^2}}W(x,m)dx\ket{n}\bra{n}
% \end{eqnarray}
% Des relations suivantes
where the following relations
\begin{equation}
z =re^{i\theta},
% \; \bar{z}=re^{-i\theta}\;\text{and}
\;\frac{d^2z}{\pi}=\frac{rdrd\theta}{\pi} 
\;\text{and}\; |z|^2 = x
\end{equation}
are used.
%(\ref{mittag-lefr0084}), \ref{mittag-lefr0086} et \ref{prelimi0087},
% {\color{blue}
% nous avons:
% \begin{eqnarray}
% &&\int_{\mathbb{C}}\frac{d^2z}{\pi}\ket{z;m}\bra{z;m}W(|z|^2,m)\cr
% &=&\sum_{n=0}^{\infty}\sum_{k=0}^{\infty}\int_{0}^{2\pi}\int_{0}^{\infty}\frac{rdrd\theta}{\pi}[N(r^2,m)]^{-1}\frac{e^{i(n-k)\theta}r^{n+k}}{\sqrt{\frac{\Gamma(n+1)(2m+\mu)_n}{[(-m-n-\nu)_n]^2}\frac{\Gamma(k+1)(2m+\mu)_k}{[(-m-k-\nu)_k]^2}}}W(r^2,m)\ket{n}\bra{k}\cr
% &=&\sum_{n=0}^{\infty}\sum_{k=0}^{\infty}\int_{0}^{2\pi}\frac{e^{i(n-k)\theta}}{\pi}d\theta\int_{0}^{\infty}[N(r^2,m)]^{-1}\frac{r^{n+k}}{\sqrt{\frac{\Gamma(n+1)(2m+\mu)_n}{[(-m-n-\nu)_n]^2}\frac{\Gamma(k+1)(2m+\mu)_k}{[(-m-k-\nu)_k]^2}}}rdr\,W(r^2,m)\ket{n}\bra{k}\cr
% &=&\sum_{n=0}^{\infty}\sum_{k=0}^{\infty}2\delta_{n,k}\int_{0}^{\infty}[N(r^2,m)]^{-1}\frac{r^{n+k}}{\sqrt{\frac{\Gamma(n+1)(2m+\mu)_n}{[(-m-n-\nu)_n]^2}\frac{\Gamma(p+1)(2m+\mu)_k}{[(-m-k-\nu)_k]^2}}}rdr\,W(r^2,m)\ket{n}\bra{k}\nonumber
% \end{eqnarray}
% Pour $n=k$, on a:
% \medskip
% \begin{equation}
% \int_{\mathbb{C}}\frac{d^2z}{\pi}\ket{z;m}\bra{z;m}W(|z|^2,m)
% &=&\sum_{n=0}^{\infty}\int_{0}^{\infty}[N(r^2,m)]^{-1}\frac{r^{2n}}{\frac{\Gamma(n+1)(2m+\mu)_n}{[(-m-n-\nu)_n]^2}}2rdr \,W(r^2,m)\ket{n}\bra{n}\cr
% \int_{\mathbb{C}}\frac{d^2z}{\pi}\ket{z;m}\bra{z;m}W(|z|^2,m)
% =\sum_{n=0}^{\infty}\int_{0}^{\infty}[N(x,m)]^{-1}\frac{x^{n}}{\frac{\Gamma(n+1)(2m+\mu)_n}{[(-m-n-\nu)_n]^2}}W(x,m)dx\ket{n}\bra{n}.%=\sum_{n=0}^{\infty}\ket{n}\bra{n}.
% \end{equation}
% }
The second member in the  last line of (\ref{resolv001}) leads to
\begin{eqnarray}{\label{houn-sodo0047}}
&&\int_{0}^{\infty}[N(x,m)]^{-1}\frac{x^{n}}{\frac{\Gamma(n+1)(2m+2\nu)_n}{[(-m-n-\nu)_n]^2}}W(x,m)dx\cr
&=&
%\int_{0}^{\infty}[N(x,m)]^{-1}\frac{\left[\frac{\Gamma(-m-\nu)}{\Gamma(-m-n-\nu)}\right]^2}{\frac{\Gamma(n+1)\Gamma(2m+\mu+n)}{\Gamma(2m+\mu)}}W(x,m)x^{n}dx&=&1\cr
\int_{0}^{\infty}\frac{[\Gamma(-m-\nu)]^2\Gamma(2m+2\nu)[N(x,m)]^{-1}W(x,m)}{[\Gamma(-m-n-\nu)]^2\Gamma(n+1)\Gamma(2m+2\nu+n)}x^{n}dx\cr
&=&1.
\end{eqnarray}
Then, setting 
{\small
\begin{eqnarray}
n&=&s-1,\cr
g^{(m)}(x)&=&[\Gamma(-m-\nu)]^2\Gamma(2m+2\nu)[N(x,m)]^{-1}W(x,m)\nonumber
\end{eqnarray}
}
and by use of the Meijer's G-functions and the Mellin inversion theorem \cite{marichev,prudnikov-brychkov-marichev}
\begin{eqnarray}
&&\int_{0}^{\infty}dx x^{s-1}
{G_{p,q}^{m,n}\left(\left. \alpha x\right| 
\begin{array}{rcl} a_{1}, \dots, a_{n};a_{n+1}, \dots, a_{p}\\b_{1},\dots,b_{m};b_{m+1},\dots,b_{q}& \end{array}\right)} \cr
&=& \frac{1}{\alpha^s}\frac{\displaystyle\prod_{j=1}^{m}\Gamma(b_j+s)}{\displaystyle\prod_{j=m+1}^{q}\Gamma(1-b_j-s)} \frac{\displaystyle\prod_{j=1}^{n}\Gamma(1-a_j-s)}{\displaystyle\prod_{j=n+1}^{p}\Gamma(a_j+s)},
\end{eqnarray}
the Eq.(\ref{houn-sodo0047}) leads to
\beq
\int_{0}^{\infty}\frac{g^{(m)}(x)}{[\Gamma(-m-s+1-\nu)]^2\Gamma(s)\Gamma(2m+2\nu+s-1)}x^{s-1}dx
=1,\nonumber
\eeq
implying 
\beq
\int_{0}^{\infty}g^{(m)}(x)x^{s-1}dx\
&=&[\Gamma(1-\nu-m-s)]^2\Gamma(s) \Gamma(2m+2\nu+s-1).
\eeq
Using the connection between  the hypergeometric functions and the  Meijer's G-functions \cite{mathai,luke}, it follows:
% De la définition des fonctions hypergéométriques généralisées et les Meijer-$G$ (voir la relation (\ref{preli0045}) au Chapitre 2), nous obtenons les relations
\begin{equation*}
g^{(m)}(x)=\MeijerG{2,0}{2,2}{m+\nu;\,\,\,m+\nu}{0;\,\,\,2m+2\nu-1}{x}, 
\end{equation*}
{\small
\beq
[\Gamma(-m-\nu)]^2\Gamma(2m+2\nu)[N(x,m)]^{-1}W(x,m)
=\MeijerG{2,0}{2,2}{m+\nu;\,\,\,m+\nu}{0;\,\,\,2m+2\nu-1}{x},\nonumber
\eeq
such that we get 
\beq
W(x,m) &=& \frac{N(x,m)}{[\Gamma(-m-\nu)]^2\Gamma(2m+2\nu)}\MeijerG{2,0}{2,2}{m+\nu;\,\,\,m+\nu}{0;\,\,\,2m+2\nu-1}{x}.
\nonumber
\eeq
Thereby, by replacing $x$ by $|z|^2$, it comes
{\small
\beq
W(|z|^2,m)&=&\frac{\MeijerG{2,0}{2,2}{m+\nu;\,\,\,m+\nu}{0;\,\,\,2m+2\nu-1}{|z|^2}}{[\Gamma(-m-\nu)\Gamma(-m-n-\nu)]^2}
\MeijerG{1,2}{2,2}{1+m+n+\nu,\,\,\,\,\,1+m+n+\nu}{0,\quad 1-2m-2\nu}{-|z|^2}.
\eeq
$\hfill{\square}$

{\bf  Proof of Proposition \ref{prop03}}

The proof of (i) and (ii) can be easily obtained. Indeed,  after direct calculations, we arrive at 
\beq
\overline{K(z,z')}&=& \frac{1}{\left[\MeijerG{1,2}{2,2}{1+m+n+\nu,\,\,\,\,\,1+m+n+\nu}{0,\,\,\,\,\,\,\,\,\,\,\,\,\,\,\,1-2m-2\nu}{-|z'|^2}\right]^{\frac{1}{2}}}\frac{\MeijerG{1,2}{2,2}{1+m+n+\nu,\,\,\,\,\,1+m+n+\nu}{0,\,\,\,\,\,\,\,\,\,\,\,\,\,\,\,1-2m-2\nu}{-z'{\bar{z}}}}{\left[\MeijerG{1,2}{2,2}{1+m+n+\nu,\,\,\,\,\,1+m+n+\nu}{0,\,\,\,\,\,\,\,\,\,\,\,\,\,\,\,1-2m-2\nu}{-|z|^2}\right]^{\frac{1}{2}}}\cr
&=& K(z',z)
\eeq
and
\beq
K(z,z)=1>0.
\eeq
We start the proof of the idempotence property by first using the expression of the reproducing kernel (\ref{kern000}), and then writing:
\begin{eqnarray}
\int_{\mathbb{C}}K(z,z")K(z",z)W(|z"|^2)\frac{d^2z"}{\pi}
%  &=&
% \int_{\mathbb{C}}\frac{\MeijerG{1,2}{2,2}{1+m+n+\nu,\,\,\,\,\,1+m+n+\nu}{0,\,\,\,\,\,\,\,\,\,\,\,\,\,\,\,1-2m-\mu}{-z{\bar{z"}}}}{\left[\MeijerG{1,2}{2,2}{1+m+n+\nu,\,\,\,\,\,1+m+n+\nu}{0,\,\,\,\,\,\,\,\,\,\,\,\,\,\,\,1-2m-\mu}{-|z"|^2}\MeijerG{1,2}{2,2}{1+m+n+\nu,\,\,\,\,\,1+m+n+\nu}{0,\,\,\,\,\,\,\,\,\,\,\,\,\,\,\,1-2m-\mu}{-|z|^2}\right]^{\frac{1}{2}}}\cr
% &&\times\frac{\MeijerG{1,2}{2,2}{1+m+n+\nu,\,\,\,\,\,1+m+n+\nu}{0,\,\,\,\,\,\,\,\,\,\,\,\,\,\,\,1-2m-\mu}{-z"{\bar{z'}}}}{\left[\MeijerG{1,2}{2,2}{1+m+n+\nu,\,\,\,\,\,1+m+n+\nu}{0,\,\,\,\,\,\,\,\,\,\,\,\,\,\,\,1-2m-\mu}{-|z"|^2}\MeijerG{1,2}{2,2}{1+m+n+\nu,\,\,\,\,\,1+m+n+\nu}{0,\,\,\,\,\,\,\,\,\,\,\,\,\,\,\,1-2m-\mu}{-|z'|^2}\right]^{\frac{1}{2}}}\cr
% &&\times\frac{\MeijerG{2,0}{2,2}{m+\nu;\,\,\,m+\nu}{0;\,\,\,2m+\mu-1}{|z"|^2}\MeijerG{1,2}{2,2}{1+m+n+\nu,\,\,\,\,\,1+m+n+\nu}{0,\quad 1-2m-\mu}{-|z"|^2}}{[\Gamma(-m-\nu)\Gamma(-m-n-\nu)]^2}\frac{d^2z"}{\pi}\cr
&=&\int_{\mathbb{C}}\MeijerG{2,0}{2,2}{m+\nu;\,\,\,m+\nu}{0;\,\,\,2m+2\nu-1}{|z"|^2}\frac{\MeijerG{1,2}{2,2}{1+m+n+\nu,\,\,\,\,\,1+m+n+\nu}{0,\,\,\,\,\,\,\,\,\,\,\,\,\,\,\,1-2m-2\nu}{-z{\bar{z"}}}}{\left[\MeijerG{1,2}{2,2}{1+m+n+\nu,\,\,\,\,\,1+m+n+\nu}{0,\,\,\,\,\,\,\,\,\,\,\,\,\,\,\,1-2m-2\nu}{-|z|^2}\right]^{\frac{1}{2}}}\cr
&&\times \frac{1}{[\Gamma(-m-\nu)\Gamma(-m-n-\nu)]^2} \cr
&&\times\frac{\MeijerG{1,2}{2,2}{1+m+n+\nu,\,\,\,\,\,1+m+n+\nu}{0,\,\,\,\,\,\,\,\,\,\,\,\,\,\,\,1-2m-2\nu}{-z"{\bar{z'}}}}{\left[\MeijerG{1,2}{2,2}{1+m+n+\nu,\,\,\,\,\,1+m+n+\nu}{0,\,\,\,\,\,\,\,\,\,\,\,\,\,\,\,1-2m-2\nu}{-|z'|^2}\right]^{\frac{1}{2}}} \frac{d^2z"}{\pi}.
\end{eqnarray}
Let
\beq
\mathfrak{S}(z,z')&=&\MeijerG{1,2}{2,2}{1+m+n+\nu,\,\,\,\,\,1+m+n+\nu}{0,\,\,\,\,\,\,\,\,\,\,\,\,\,\,\,1-2m-2\nu}{-z{\bar{z"}}}\MeijerG{1,2}{2,2}{1+m+n+\nu,\,\,\,\,\,1+m+n+\nu}{0,\,\,\,\,\,\,\,\,\,\,\,\,\,\,\,1-2m-2\nu}{-z"{\bar{z'}}}
\eeq
and
\beq
\mathfrak X (z")=\int_{\mathbb{C}}\frac{\MeijerG{2,0}{2,2}{m+\nu;\,\,\,m+\nu}{0;\,\,\,2m+2\nu-1}{|z"|^2}}{[\Gamma(-m-\nu)\Gamma(-m-n-\nu)]^2}\mathfrak S(z,z')\frac{d^2z"}{\pi}.
\eeq
From 
\beq
\mathfrak S(z,z')
% &=&\sum_{n=0}^{\infty}\frac{[(-m-n-\nu)_n]^2[\Gamma(-m-n-\nu)]^2}{(2m+\mu)_n\Gamma(2m+\mu)}\frac{(\bar{z"}z)^n}{\Gamma(n+1)}\cr
% &&\times\sum_{k=0}^{\infty}\frac{[(-m-k-\nu)_k]^2[\Gamma(-m-k-\nu)]^2}{(2m+\mu)_k\Gamma(2m+\mu)}\frac{(\bar{z'}z")^k}{\Gamma(k+1)}\cr
% &=&\sum_{n=0}^{\infty}\frac{[\Gamma(-m-\nu)]^2}{\Gamma(2m+\mu+n)}\frac{(\bar{z"}z)^n}{\Gamma(n+1)}\sum_{k=0}^{\infty}\frac{[\Gamma(-m-\nu)]^2}{\Gamma(2m+\mu+k)}\frac{(\bar{z'}z")^k}{\Gamma(k+1)}\cr
=\sum_{n=0}^{\infty}\sum_{k=0}^{\infty}\frac{[\Gamma(-m-\nu)]^4}{\Gamma(2m+2\nu+n)\Gamma(2m+2\nu+k)}\frac{(\bar{z"}z)^n(\bar{z'}z")^k}{\Gamma(n+1)\Gamma(k+1)}
\eeq
we get 
\beq
\mathfrak X(z")
&=&\sum_{n=0}^{\infty}\sum_{k=0}^{\infty}\frac{[\Gamma(-m-\nu)]^4}{\Gamma(2m+2\nu+n)\Gamma(2m+2\nu+k)}\frac{(z)^n(\bar{z'})^k}{\Gamma(n+1)\Gamma(k+1)}\cr
&&\times\int_{\mathbb{C}}\frac{\MeijerG{2,0}{2,2}{m+\nu;\,\,\,m+\nu}{0;\,\,\,2m+2\nu-1}{|z"|^2}}{[\Gamma(-m-\nu)\Gamma(-m-n-\nu)]^2}(\bar{z"})^n(z")^k\frac{d^2z"}{\pi}\cr
&=&\MeijerG{1,2}{2,2}{1+m+n+\nu,\,\,\,\,\,1+m+n+\nu}{0,\,\,\,\,\,\,\,\,\,\,\,\,\,\,\,1-2m-2\nu}{-z{\bar{z'}}}.
\eeq
Thereby
\beq
\int_{\mathbb{C}}K(z,z")K(z",z)W(|z"|^2)\frac{d^2z"}{\pi}
&=& \frac{1}{\left[\MeijerG{1,2}{2,2}{1+m+n+\nu,\,\,\,\,\,1+m+n+\nu}{0,\,\,\,\,\,\,\,\,\,\,\,\,\,\,\,1-2m-2\nu}{-|z|^2}\right]^{\frac{1}{2}}}\cr
&&\times
\frac{\MeijerG{1,2}{2,2}{1+m+n+\nu,\,\,\,\,\,1+m+n+\nu}{0,\,\,\,\,\,\,\,\,\,\,\,\,\,\,\,1-2m-2\nu}{-z{\bar{z'}}}}{\left[\MeijerG{1,2}{2,2}{1+m+n+\nu,\,\,\,\,\,1+m+n+\nu}{0,\,\,\,\,\,\,\,\,\,\,\,\,\,\,\,1-2m-2\nu}{-|z'|^2}\right]^{\frac{1}{2}}}\cr
&=&K(z,z').
\eeq

$\hfill{\square}$

{\bf Proof of Proposition \ref{prop02}}

From the relations  (\ref{houn-sodo0042}), the expectation of a given observable $A$ in the basis of the CSs  $\ket{z;m}$ is obtained as follows:
\begin{eqnarray}
\bra{z;m}{A}\ket{z;m} = \braket{{A}}_{z,m} 
&=&
\left[\frac{\Gamma(2m+2\nu)}{[\Gamma(-m-k-\nu)]^2}\MeijerG{1,2}{2,2}{1+m+k+\nu,\,\,\,\,\,1+m+k+\nu}{0,\,\,\,\,\,\,\,\,\,\,\,\,\,\,\,1-2m-2\nu}{-|z|^2}\right.\cr
&&\left.\times \frac{\Gamma(2m+2\nu)}{[\Gamma(-m-n-\nu)]^2}\MeijerG{1,2}{2,2}{1+m+n+\nu,\,\,\,\,\,1+m+n+\nu}{0,\,\,\,\,\,\,\,\,\,\,\,\,\,\,\,1-2m-2\nu}{-|z|^2}\right]^{-\frac{1}{2}}\cr
&&\times\sum_{n,k=0}^{\infty}\frac{z^n \bar{z}^k}{\sqrt{\frac{\Gamma(n+1)(2m+2\nu)_n}{[(-m-n-\nu)_n]^2}\frac{\Gamma(k+1)(2m+2\nu)_k}{[(-m-k-\nu)_k]^2}}}\bra{k}{A}\ket{n}.
\end{eqnarray}
Thereby, if $n=k$, it comes 
\begin{eqnarray}
 \braket{{A}}_{z,m}
% &=&\left[\frac{\Gamma(2m+\mu)}{[\Gamma(-m-n-\nu)]^2}\MeijerG{1,2}{2,2}{1+m+n+\nu,\,\,\,\,\,1+m+n+\nu}{0,\,\,\,\,\,\,\,\,\,\,\,\,\,\,\,1-2m-\mu}{-|z|^2}\right]^{-1}\cr
% &&\times\sum_{n=0}^{\infty}\frac{z^{2n}} {\frac{\Gamma(n+1)(2m+\mu)_n}{[(-m-n-\nu)_n]^2}}\bra{n}A\ket{n}\cr
% &=&
&=&[N(|z|^2,m)]^{-1}\sum_{n=0}^{\infty}\frac{[(-m-n-\nu)_n]^2}{(2m+2\nu)_n}\frac{z^{2n}} {\Gamma(n+1)}\bra{n}A\ket{n}.
\end{eqnarray} 
For the number operator  ${N}$, we have in the Fock basis $\{\ket{n}\}_{n=0}^{\infty}$: 
$\bra{n}N^i\ket{n}=n^i$ and $\bra{n'}N^i\ket{n}=0$.
Using the ansatz \cite{popov1}, set
\beq
S_i&=&\sum_{n=0}^{\infty}\frac{[(-m-n-\nu)_n]^2}{(2m+2\nu)_n}\frac{x^n}{\Gamma(n+1)}n^i\;\text{with}\; x=|z|^2.
\eeq
% For the particular case  $i=0$, we have
% \begin{eqnarray}
% S_0&=&\sum_{n=0}^{\infty}\frac{[(-m-n-\nu)_n]^2}{(2m+\mu)_n}\frac{x^n}{\Gamma(n+1)}\cr
% S_0&=&\frac{\Gamma(2m+\mu)}{[\Gamma(-m-n-\nu)]^2}\MeijerG{1,2}{2,2}{1+m+n+\nu,\,\,\,\,1+m+n+\nu}{0,\,\,\,\,1-2m-\mu}{-x}.
% \end{eqnarray}
For  $i>0$, we obtain the following relation for  $n^i$
\begin{eqnarray}
n^i&=&\sum_{l=1}^{i}C_l^{(i)}\frac{n!}{(n-l)!}=\sum_{l=0}^{i}(-1)^lC_l^{(i)}\frac{n!}{(n-l)!}.
\end{eqnarray}
Then, we get 
\begin{eqnarray}
S_i =\left(x\frac{d}{dx}\right)^iS_0=\sum_{l=1}^{i}C_l^{(n)}x^l\left(\frac{d}{dx}\right)^lS_0, 
% \cr
% &=&\sum_{l=1}^{i}C_l^{(n)}x^l\left(\frac{d}{dx}\right)^l\frac{\Gamma(2m+\mu)}{[\Gamma(-m-n-\nu)]^2}\MeijerG{1,2}{2,2}{1+m+n+\nu,\,\,\,\,1+m+n+\nu}{0,\,\,\,\,\,\,\,\,1-2m-\mu}{-x},
\end{eqnarray}
where 
\beq
S_0  = \frac{\Gamma(2m+2\nu)}{[\Gamma(-m-n-\nu)]^2}\MeijerG{1,2}{2,2}{1+m+n+\nu,\,\,\,\,1+m+n+\nu}{0,\,\,\,\,1-2m-2\nu}{-x}
\eeq
such that, by applying the $nth$ derivative of the 
% en appliquant la dérivée $n^{\text{ième}}$
% derivative of the 
% des fonctions
 Meijer's G-functions, we get:
\begin{eqnarray}\label{sumderv000}
S_i&=&\sum_{l=1}^{i}(-1)^lC_l^{(n)}\frac{\Gamma(2m+2\nu)}{[\Gamma(-m-n-\nu)]^2}\MeijerG{2,2}{3,3}{1+m+n+\nu,\,\,\,\,1+m+n+\nu,\,\,\,\,0}{0,\,\,\,\,\,\,\,\,l,\,\,\,\,\,\,\,\,1-2m-2\nu}{-x}\cr
&=&\sum_{l=1}^{i}(-1)^lC_l^{(n)}\frac{\Gamma(2m+2\nu)}{[\Gamma(-m-n-\nu)]^2}\MeijerG{1,2}{2,2}{1+m+n+\nu,\,\,\,\,1+m+n+\nu}{l,\,\,\,\,\,\,\,\,1-2m-2\nu}{-x}.
\end{eqnarray}
From (\ref{sumderv000}), the expectations of the operators $N$ and $N^2$ are:
\beq
\braket{N}_{z,m}
% &=&[N(|z|^2,m)]^{-1}\sum_{n=0}^{\infty}\frac{[\Gamma(-m-n-\nu)]^2}{\Gamma(2m+\mu)}\frac{|z|^{2n}}{n!}n\cr
&=&[N(|z|^2,m)]^{-1}S_1 = -C_1^{(n)}\frac{\MeijerG{1,2}{2,2}{1+m+n+\nu,\,\,\,\,1+m+n+\nu}{1,\,\,\,\,\,\,\,\,1-2m-2\nu}{-|z|^2}}{\MeijerG{1,2}{2,2}{1+m+n+\nu,\,\,\,\,1+m+n+\nu}{0,\,\,\,\,\,\,\,\,1-2m-2\nu}{-|z|^2}},
\eeq
and 
\begin{eqnarray}
\braket{{N}^{2}}_{z,m}
% &=&[N(|z|^2,m)]^{-1}\sum_{n=0}^{\infty}\frac{[\Gamma(-m-n-\nu)]^2}{\Gamma(2m+\mu)}\frac{|z|^{2n}}{n!}n^2\cr
&=&[N(|z|^2,m)]^{-1}S_2\cr &=&\frac{1}{\MeijerG{1,2}{2,2}{1+m+n+\nu,\,\,\,\,1+m+n+\nu}{0,\,\,\,\,\,\,\,\,1-2m-2\nu}{-|z|^2}}\left[-C_1^{(n)}\MeijerG{1,2}{2,2}{1+m+n+\nu,\,\,\,\,1+m+n+\nu}{1,\,\,\,\,\,\,\,\,1-2m-2\nu}{-|z|^2} \right.\cr && \left. + C_2^{(n)}\MeijerG{1,2}{2,2}{1+m+n+\nu,\,\,\,\,1+m+n+\nu}{2,\,\,\,\,\,\,\,\,1-2m-2\nu}{-|z|^2}\right].
\end{eqnarray}

$\hfill{\square}$
{
	%\color{blue}
\section*{Acknowledgment}
%This work is supported by TWAS Research Grant RGA 
%No. 17-542 RG / MATHS / AF / AC \_G -FR3240300147. 
The ICMPA-UNESCO Chair is in partnership with Daniel Iagolnitzer Foundation (DIF), 
France, supporting the development of mathematical physics in Africa.
}

\end{document}